\documentclass[11pt,onecolumn]{article}

\usepackage{amsfonts,epsfig,graphicx,subfigure}
\usepackage{amsmath,amssymb,amsthm}

\usepackage{epic,eepic}

\usepackage{epsf}
\usepackage{fancyheadings}
\usepackage{graphics}
\usepackage{amsfonts}
\usepackage{amsmath}
\usepackage{amssymb}
\usepackage{epic}
\usepackage{eepic}
\usepackage{eepicemu}
\usepackage{psfrag}
\usepackage{setspace}

\usepackage{arxivmacros}

\newcommand{\kval}{\ensuremath{k}}

\newcommand{\myhalf}{\ensuremath{\frac{1}{2}}}

\newcommand{\numcheck}{\ensuremath{m}}
\newcommand{\numbit}{\ensuremath{n}}

\newcommand{\myparagraph}[1]{{\bf{#1}}}

\newcommand{\midbit}{\ensuremath{m}}

\newcommand{\myber}{\ensuremath{\operatorname{Ber}}}

\newcommand{\ratewz}{\ensuremath{R_{\operatorname{WZ}}}}

\newcommand{\rateie}{\ensuremath{R_{\operatorname{IE}}}}
\newcommand{\bernoise}{\ensuremath{\delta}}

\newcommand{\topbit}{\ensuremath{n}}
\newcommand{\lowbit}{\ensuremath{k}}
\newcommand{\mydefn}{\ensuremath{: \, =}}
\newcommand{\estim}[1]{\ensuremath{\widehat{#1}}}

\newcommand{\iew}{\ensuremath{w}}
\newcommand{\channoise}{\ensuremath{p}}

\newcommand{\uce}{\ensuremath{\operatorname{u.c.e.}}}
\newcommand{\lce}{\ensuremath{\operatorname{l.c.e.}}}

\newcommand{\rval}{\ensuremath{t}}
\newcommand{\MomGen}[1]{\ensuremath{\mathbb{M}_{#1}}}
\newcommand{\lamstar}{\ensuremath{\lambda^*}}
\newcommand{\tmpq}{\ensuremath{u}}

\newcommand{\rateeff}{\ensuremath{R_{\operatorname{trans}}}}
\newcommand{\KeyFunc}{\ensuremath{F}}
\newcommand{\InterFunc}{\ensuremath{G}}

\newcommand{\MyWtEnum}{\ensuremath{\mathbb{W}}}

\newcommand{\Parmat}{\ensuremath{H}}
\newcommand{\Genmat}{\ensuremath{G}}

\newcommand{\rateldgm}{\ensuremath{R_{\Genmat}}}
\newcommand{\rateldpc}{\ensuremath{R_{\Parmat}}}
\newcommand{\rateldpcone}{\ensuremath{R_{\Parmat_1}}}

\newcommand{\ratetot}{\ensuremath{R}}
\newcommand{\ratecom}{\ensuremath{\ratetot}}

\newcommand{\rtotvar}{\ensuremath{W}}
\newcommand{\rvaradd}{\ensuremath{U}}
\newcommand{\rvarplain}{\ensuremath{V}}
\newcommand{\rcountvar}{\ensuremath{T}}
\newcommand{\indber}[1]{\ensuremath{\inducedDmin{#1}}}
\newcommand{\IndBer}[1]{\indber{#1}}

\newcommand{\lowtay}{\ensuremath{\mu}}

\newcommand{\ErrFun}{\ensuremath{L}}
\newcommand{\ErrFunTil}{\ensuremath{\wtil{\ErrFun}}}

\newcommand{\ldpcthresh}{\ensuremath{\nu^*}}

\newcommand{\zeroes}{\ensuremath{0^\topbit}}

\newcommand{\Numc}{\ensuremath{T_\topbit}}
\newcommand{\mprob}{\ensuremath{\mathbb{P}}}
\newcommand{\cwind}[1]{\ensuremath{Z^{#1}}}
\newcommand{\cw}[1]{\ensuremath{X^{#1}}}

\newcommand{\Ysca}{\ensuremath{Y}}

\newcommand{\Ssca}{\ensuremath{S}}

\newcommand{\Svar}{\ensuremath{S}}

\newcommand{\Qnoise}{\ensuremath{E}}

\newcommand{\codebit}{\ensuremath{x}}
\newcommand{\infobit}{\ensuremath{y}}

\newcommand{\Codebit}{\ensuremath{X}}

\newcommand{\Recbit}{\ensuremath{V}}

\newcommand{\numcodebit}{\ensuremath{n}}
\newcommand{\numinfobit}{\ensuremath{m}}
\newcommand{\mycode}{\ensuremath{\mathbb{C}}}

\newcommand{\acoeff}{\ensuremath{a}}
\newcommand{\bcoeff}{\ensuremath{b}}
\newcommand{\ccoeff}{\ensuremath{c}}

\newcommand{\mywei}{\ensuremath{w}}
\newcommand{\myweibar}{\ensuremath{\widetilde{\mywei}}}
\newcommand{\Qprob}{\ensuremath{\mathbb{Q}}}
\newcommand{\AvWtEnum}[1]{\ensuremath{\mathbb{A}_{#1}}}
\newcommand{\BouWtEnum}{\ensuremath{B}}
\newcommand{\AsympWtEnum}{\ensuremath{B}}

\newcommand{\delfun}[1]{\ensuremath{\delta^*(#1; \topdeg)}}
\newcommand{\tmpvar}{\ensuremath{t}}

\newcommand{\Tmpfun}{\ensuremath{g}}

\newcommand{\myrate}{\ensuremath{R}}

\newcommand{\lowcdeg}{\ensuremath{d'_c}}

\newcommand{\mytvar}{\ensuremath{t}}

\newcommand{\mylam}{\ensuremath{\lambda}}

\newcommand{\ProofFun}{\ensuremath{K}}

\newcommand{\epsup}{\ensuremath{\epsilon_2}}
\newcommand{\epslow}{\ensuremath{\epsilon_1}}

\newcommand{\Wnoise}{\ensuremath{W}}

\newcommand{\myeps}{\ensuremath{\epsilon_\topbit}}

\newcommand{\Wzside}{\ensuremath{Z}}

\newcommand{\Kset}{\ensuremath{K}}
\newcommand{\mycodepar}{\ensuremath{\mycode(\Genmat, \Parmat_1)}}

\newcommand{\msca}{\ensuremath{\genericS{m}}}

\newcommand{\Gpmess}{\ensuremath{\msca}}
\newcommand{\gpmsca}{\ensuremath{\msca}}

\newcommand{\Chaninput}{\ensuremath{V}}
\newcommand{\Chanoutput}{\ensuremath{Z}}
\newcommand{\Shost}{\ensuremath{S}}

\newcommand{\mycodegp}{\ensuremath{\mycode}}

\newcommand{\ustar}{\ensuremath{u^*}}

\newcommand{\ratesha}{\ensuremath{R_{\operatorname{Sha}}}}

\newcommand{\Complexnumc}{\ensuremath{T_\topbit(\Ssca, \mycode;
\distor)}}

\newcommand{\rvdistcom}{\ensuremath{d_\topbit(\Ssca, \mycode)}}

\newcommand{\mycodetop}{\ensuremath{\mycode_1}}
\newcommand{\mycodebot}{\ensuremath{\mycode_2}}
\newcommand{\mycodebarbot}{\ensuremath{\bar{\mycode}_2}}
\newcommand{\mycodebar}{\ensuremath{\bar{\mycode}}}

\newcommand{\mycodebarldpc}{\mycodebarbot}

\newcommand{\Sspace}{\ensuremath{\mathcal{S}}}

\newcommand{\mycond}[2]{\ensuremath{\mprob(#1 \mid #2)}}
\newcommand{\cond}{\mycond}
\newcommand{\xml}{\ensuremath{\widehat{x}}}

\newcommand{\shat}{\ensuremath{\widehat{S}}}

\newcommand{\real}{\ensuremath{\mathbb{R}}}

\begin{document}
\begin{center}

{{\LARGE \bf{Low-density graph codes that are optimal for
source/channel coding and binning}}}

\begin{center}
\begin{tabular}{ccc}
Martin J. Wainwright & & Emin Martinian\\
Dept. of Statistics, and  & &   Tilda Consulting, Inc. \\
Dept. of Electrical Engineering and Computer Sciences  & & Arlington, MA \\
University of California, Berkeley & &  \texttt{emin@alum.mit.edu} \\
 \texttt{wainwrig@\{eecs,stat\}.berkeley.edu} & & 
\end{tabular}
\end{center}

\vspace*{.5in}

Technical Report 730, \\
Department of Statistics, UC Berkeley, \\
April 2007
\end{center}

\begin{abstract}
We describe and analyze the joint source/channel coding properties of
a class of sparse graphical codes based on compounding a low-density
generator matrix (LDGM) code with a low-density parity check (LDPC)
code.  Our first pair of theorems establish that there exist codes
from this ensemble, with all degrees remaining bounded independently
of block length, that are simultaneously optimal as both source and
channel codes when encoding and decoding are performed optimally.
More precisely, in the context of lossy compression, we prove that
finite degree constructions can achieve any pair $(\myrate, \distor)$
on the rate-distortion curve of the binary symmetric source.  In the
context of channel coding, we prove that finite degree codes can
achieve any pair $(C, \channoise)$ on the capacity-noise curve of the
binary symmetric channel.  Next, we show that our compound
construction has a nested structure that can be exploited to achieve
the Wyner-Ziv bound for source coding with side information (SCSI), as
well as the Gelfand-Pinsker bound for channel coding with side
information (CCSI).  Although the current results are based on optimal
encoding and decoding, the proposed graphical codes have sparse
structure and high girth that renders them well-suited to
message-passing and other efficient decoding procedures.
\end{abstract}

\noindent {\bf Keywords:} Graphical codes; low-density parity check
code (LDPC); low-density generator matrix code (LDGM); weight
enumerator; source coding; channel coding; Wyner-Ziv problem;
Gelfand-Pinsker problem; coding with side information; information
embedding; distributed source coding.

\begin{onehalfspace}

\section{Introduction}

Over the past decade, codes based on graphical constructions,
including turbo codes~\cite{Berroux96} and low-density parity check
(LDPC) codes~\cite{Gallager63}, have proven extremely successful for
channel coding problems.  The sparse graphical nature of these codes
makes them very well-suited to decoding using efficient
message-passing algorithms, such as the sum-product and max-product
algorithms.  The asymptotic behavior of iterative decoding on graphs
with high girth is well-characterized by the density evolution
method~\cite{Luby01,Richardson01a}, which yields a useful design
principle for choosing degree distributions.  Overall, suitably
designed LDPC codes yield excellent practical performance under
iterative message-passing, frequently very close to Shannon
limits~\cite{CFRU}.

However, many other communication problems involve aspects of lossy
source coding, either alone or in conjunction with channel coding, the
latter case corresponding to joint source-channel coding
problems. Well-known examples include lossy source coding with side
information (one variant corresponding to the Wyner-Ziv
problem~\cite{WynerZiv76}), and channel coding with side information
(one variant being the Gelfand-Pinsker problem~\cite{Gelfand80}).  The
information-theoretic schemes achieving the optimal rates for coding
with side information involve delicate combinations of source and
channel coding.  For problems of this nature---in contrast to the case
of pure channel coding---the use of sparse graphical codes and
message-passing algorithm is not nearly as well understood.  With
this perspective in mind, the focus of this paper is the design and
analysis sparse graphical codes for lossy source coding, as well as
joint source/channel coding problems.  Our main contribution is to
exhibit classes of graphical codes, with all degrees remaining bounded
independently of the blocklength, that simultaneously achieve the
information-theoretic bounds for both source and channel coding under
optimal encoding and decoding.

\subsection{Previous and ongoing work}

A variety of code architectures have been suggested for lossy
compression and related problems in source/channel coding.  One
standard approach to lossy compression is via trellis-code
quantization (TCQ)~\cite{Marcellin90}.  The advantage of trellis
constructions is that exact encoding and decoding can be performed
using the max-product or Viterbi algorithm~\cite{Loeliger04}, with
complexity that grows linearly in the trellis length but exponentially
in the constraint length.  Various researchers have exploited
trellis-based codes both for single-source and distributed
compression~\cite{Chou03,Liveris03,Pradhan03,Yang05} as well as
information embedding problems~\cite{Chou01,Erez05,Sun05}.  One
limitation of trellis-based approaches is the fact that saturating
rate-distortion bounds requires increasing the trellis constraint
length~\cite{Viterbi74}, which incurs exponential complexity (even for
the max-product or sum-product message-passing algorithms).

Other researchers have proposed and studied the use of low-density
parity check (LDPC) codes and turbo codes, which have proven extremely
successful for channel coding, in application to various types of
compression problems.  These techniques have proven particularly
successful for \emph{lossless} distributed compression, often known as
the Slepian-Wolf problem~\cite{Garcia02,Schonberg02}.  An attractive
feature is that the source encoding step can be transformed to an
equivalent noisy channel decoding problem, so that known constructions
and iterative algorithms can be leveraged.  For \emph{lossy}
compression, other work~\cite{MatYam03} shows that it is possible to
approach the binary rate-distortion bound using LDPC-like codes,
albeit with degrees that grow logarithmically with the blocklength.

A parallel line of work has studied the use of low-density generator
matrix (LDGM) codes, which correspond to the duals of LDPC codes, for
lossy compression
problems~\cite{Martinian03,WaiMan05,Ciliberti05b,Murayama03,Murayama04}.
Focusing on binary erasure quantization (a special compression problem
dual to binary erasure channel coding), Martinian and
Yedidia~\cite{Martinian03} proved that LDGM codes combined with
modified message-passing can saturate the associated rate-distortion
bound.  Various researchers have used techniques from statistical
physics, including the cavity method and replica methods, to provide
non-rigorous analyses of LDGM performance for lossy compression of
binary sources~\cite{Ciliberti05a,Ciliberti05b,Murayama03,Murayama04}.
In the limit of zero-distortion, this analysis has been made rigorous
in a sequence of
papers~\cite{Creignou03,MezRicZec02,Cocco03,Dubois03}.  Moreover, our
own recent work~\cite{MarWai06a,MarWai06c} provides rigorous upper
bounds on the effective rate-distortion function of various classes of
LDGM codes.  In terms of practical algorithms for lossy binary
compression, researchers have explored variants of the sum-product
algorithm~\cite{Murayama04} or survey propagation
algorithms~\cite{Ciliberti05a,WaiMan05} for quantizing binary sources.

\subsection{Our contributions}  Classical random coding arguments~\cite{Cover} 
show that random binary linear codes will achieve both channel
capacity and rate-distortion bounds.  The challenge addressed in this
paper is the design and analysis of codes with \emph{bounded graphical
complexity}, meaning that all degrees in a factor graph representation
of the code remain bounded independently of blocklength.  Such
sparsity is critical if there is any hope to leverage efficient
message-passing algorithms for encoding and decoding.  With this
context, the primary contribution of this paper is the analysis of
sparse graphical code ensembles in which a low-density generator
matrix (LDGM) code is compounded with a low-density parity check
(LDPC) code (see~\figref{FigCompound} for an illustration).  Related
compound constructions have been considered in previous work, but
focusing exclusively on channel
coding~\cite{Etesami06,Pfister05,Shokrollahi06}.  In contrast, this
paper focuses on communication problems in which source coding plays
an essential role, including lossy compression itself as well as joint
source/channel coding problems.  Indeed, the source coding analysis of
the compound construction requires techniques fundamentally different
from those used in channel coding analysis.  We also note that the
compound code illustrated in~\figref{FigCompound} can be applied to
more general memoryless channels and sources; however, so as to bring
the primary contribution into sharp focus, this paper focuses
exclusively on binary sources and/or binary symmetric channels.

More specifically, our first pair of theorems establish that for any
rate $\myrate \in (0,1)$, there exist codes from compound LDGM/LDPC
ensembles with all degrees remaining bounded independently of the
blocklength that achieve both the channel capacity and the
rate-distortion bound.  To the best of our knowledge, this is the
first demonstration of code families with bounded graphical complexity
that are simultaneously optimal for both source and channel coding.
Building on these results, we demonstrate that codes from our ensemble
have a naturally ``nested'' structure, in which good channel codes can
be partitioned into a collection of good source codes, and vice versa.
By exploiting this nested structure, we prove that codes from our
ensembles can achieve the information-theoretic limits for the binary
versions of both the problem of lossy source coding with side
information (SCSI, known as the Wyner-Ziv problem~\cite{WynerZiv76}),
and channel coding with side information (CCSI, known as the
Gelfand-Pinsker~\cite{Gelfand80} problem).  Although these results are
based on optimal encoding and decoding, a code drawn randomly from our
ensembles will, with high probability, have high girth and good
expansion, and hence be well-suited to message-passing and other
efficient decoding procedures.

The remainder of this paper is organized as follows.
Section~\ref{SecBackground} contains basic background material and
definitions for source and channel coding, and factor graph
representations of binary linear codes.  In
Section~\ref{SecGenConstruc}, we define the ensembles of compound
codes that are the primary focus of this paper, and state (without
proof) our main results on their source and channel coding optimality.
In Section~\ref{SecSideInformation}, we leverage these results to show
that our compound codes can achieve the information-theoretic limits
for lossy source coding with side information (SCSI), and channel
coding with side information (CCSI).  Sections~\ref{SecSourceOptimal}
and~\ref{SecChannelOptimal} are devoted to proofs that codes from the
compound ensemble are optimal for lossy source coding
(Section~\ref{SecSourceOptimal}) and channel coding
(Section~\ref{SecChannelOptimal}) respectively.  We conclude the paper
with a discussion in Section~\ref{SecDiscussion}.  Portions of this
work have previously appeared as conference
papers~\cite{MarWai06a,MarWai06b,MarWai06c}.

\section{Background}
\label{SecBackground}

In this section, we provide relevant background material on source and
channel coding, binary linear codes, as well as factor graph
representations of such codes.

\subsection{Source and channel coding}

A binary linear code $\mycode$ of block length $\numbit$ consists of
all binary strings $x \in \{0,1\}^\numbit$ satisfying a set of
$\numcheck < \numbit$ equations in modulo two arithmetic. More
precisely, given a parity check matrix $\Parmat \in \{0,1\}^{\numcheck
\times \numbit}$, the code is given by the null space
\begin{eqnarray}
\label{EqnDefnBinLinCode}
\mycode & \mydefn & \left \{ x \in \{0,1\}^\numbit \; \mid \; \Parmat
x = 0 \right \}.
\end{eqnarray}
Assuming the parity check matrix $\Parmat$ is full rank, the code
$\mycode$ consists of $2^{\numbit - \numcheck} = 2^{\numbit \myrate}$
codewords, where $\myrate = 1 - \frac{\numcheck}{\numbit}$ is the code
rate. \\

\noindent \myparagraph{Channel coding:} In the channel coding problem,
the transmitter chooses some codeword $x\in \mycode$ and transmits it
over a noisy channel, so that the receiver observes a noise-corrupted
version $Y$.  The channel behavior is modeled by a conditional
distribution $\mycond{y}{x}$ that specifies, for each transmitted
sequence $Y$, a probability distribution over possible received
sequences $\{Y = y\}$.  In many cases, the channel is memoryless,
meaning that it acts on each bit of $\mycode$ in an independent
manner, so that the channel model decomposes as $\mycond{y}{x} =
\prod_{i=1}^\numbit f_i(x_i; y_i)$ Here each function $f_i(x_i; y_i) =
\mycond{y_i}{x_i}$ is simply the conditional probability of observing
bit $y_i$ given that $x_i$ was transmitted. As a simple example, in
the binary symmetric channel (BSC), the channel flips each transmitted
bit $x_i$ with probability $p$, so that $\mycond{y_i}{x_i} = (1-p) \,
\mathbb{I}[x_i = y_i] + p \left(1-\mathbb{I}[x_i \neq y_i] \right)$,
where $\mathbb{I}(A)$ represents an indicator function of the event
$A$.  With this set-up, the goal of the receiver is to solve the
\emph{channel decoding problem}: estimate the most likely transmitted
codeword, given by $\xml \mydefn \arg \max \limits_{x \in \mycode}
\cond{y}{x}$.  The Shannon capacity~\cite{Cover} of a channel
specifies an upper bound on the rate $\myrate$ of any code for which
transmission can be asymptotically error-free.  Continuing with our
example of the BSC with flip probability $p$, the capacity is given by
$C = 1 - h(p)$, where $h(p) \mydefn -p \log_2 p - (1-p) \log_2(1-p)$
is the binary entropy function. \\

\noindent \myparagraph{Lossy source coding:} In a lossy source coding
problem, the encoder observes some source sequence $S \in
\Sspace$, corresponding to a realization of some random vector with
i.i.d. elements $S_i \sim \mprob_S$.  The idea is to compress the
source by representing each source sequence $S$ by some codeword $x
\in \mycode$.  As a particular example, one might be interested in
compressing a \emph{symmetric Bernoulli source}, consisting of binary
strings $S \in \{0,1\}^{n}$, with each element $S_i$ drawn in an
independent and identically distributed (i.i.d.) manner from a
Bernoulli distribution with parameter $p = \frac{1}{2}$.  One could
achieve a given compression rate $\myrate = \frac{m}{n}$ by mapping
each source sequence to some codeword $x \in \mycode$ from a code
containing $2^m = 2^{n \myrate}$ elements, say indexed by the binary
sequences $z \in \{0,1\}^m$.  In order to assess the quality of the
compression, we define a source decoding map $x \mapsto \shat(x)$,
which associates a source reconstruction $\shat(x)$ with each codeword
$x \in \mycode$.  Given some distortion metric $d:\Sspace \times
\Sspace \rightarrow \real_+$, the \emph{source encoding problem} is to
find the codeword with minimal distortion---namely, the optimal
encoding $\xml \mydefn \arg \min \limits_{x \in \mycode} d(\shat(x),
S)$.  Classical rate-distortion theory~\cite{Cover} specifies the
optimal trade-offs between the compression rate $\rate$ and the best
achievable average distortion $D = \Exs[d(\estim{S}, S)]$, where the
expectation is taken over the random source sequences $S$.  For
instance, to follow up on the Bernoulli compression example, if we use
the Hamming metric $d(\shat, S) = \frac{1}{n} \sum_{i=1}^{n}
|\widehat{S}_i - S_i|$ as the distortion measure, then the
rate-distortion function takes the form $R(D) = 1 - h(D)$, where $h$
is the previously defined binary entropy function. \\

\noindent We now provide definitions of ``good'' source and channel
codes that are useful for future reference.

\bdes
\label{DefnCoding}
(a) A code family is a \emph{good $\distor$-distortion binary
symmetric source code} if for any $\epsilon > 0$, there exists a code
with rate $\rate < 1 - \binent{\distor} + \epsilon$ that achieves
Hamming distortion less than or equal to $\distor$. \\
(b) A code family is a \emph{good BSC($\channoise$)-noise channel
code} if for any $\epsilon > 0$ there exists a code with rate $\rate >
1 - \binent{\channoise} - \epsilon$ with error probability less than
$\epsilon$.
\edes

\subsection{Factor graphs and graphical codes}

Both the channel decoding and source encoding problems, if viewed
naively, require searching over an exponentially large codebook (since
$|\mycode| = 2^{\numbit \myrate}$ for a code of rate $\myrate$).
Therefore, any practically useful code must have special structure
that facilitates decoding and encoding operations.  The success of a
large subclass of modern codes in use today, especially low-density
parity check (LDPC) codes~\cite{Gallager63,Richardson01b}, is based on
the sparsity of their associated factor graphs.

\newcommand{\smallhat}{\ensuremath{x}}
\begin{figure}[h]
\begin{center}
\begin{tabular}{cc}
\psfrag{#m#}{$\numcheck$} \psfrag{#n#}{$\numbit$} \psfrag{#x1#}{$y_1$}
\psfrag{#x2#}{$y_2$} \psfrag{#x3#}{$y_3$} \psfrag{#x4#}{$y_4$}
\psfrag{#x5#}{$y_5$} \psfrag{#x6#}{$y_6$} \psfrag{#x7#}{$y_7$}
\psfrag{#x8#}{$y_8$} \psfrag{#x9#}{$y_9$} \psfrag{#x10#}{$y_{10}$}
\psfrag{#x11#}{$y_{11}$} \psfrag{#x12#}{$y_{12}$} \psfrag{#c1#}{$c_1$}
\psfrag{#c2#}{$c_2$} \psfrag{#c3#}{$c_3$} \psfrag{#c4#}{$c_4$}
\psfrag{#c5#}{$c_5$} \psfrag{#c6#}{$c_6$}
\widgraph{.5\textwidth}{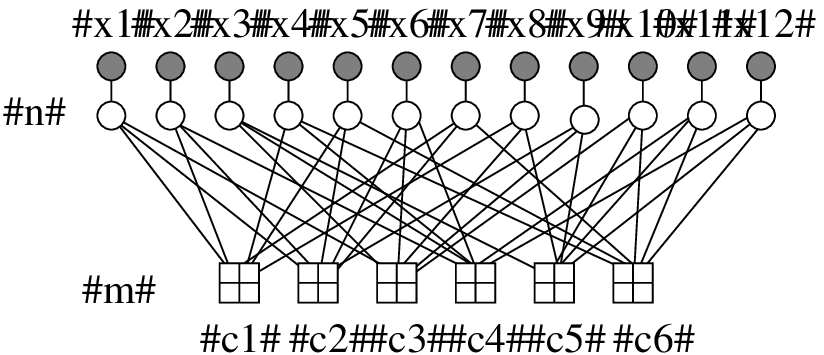} &
\psfrag{#m#}{$\numcheck$}
\psfrag{#n#}{$\numbit$}
\psfrag{#y1#}{$\smallhat1$}
\psfrag{#y2#}{$\smallhat2$}
\psfrag{#y3#}{$\smallhat3$}
\psfrag{#y4#}{$\smallhat4$}
\psfrag{#y5#}{$\smallhat5$}
\psfrag{#y6#}{$\smallhat6$}
\psfrag{#y7#}{$\smallhat7$}
\psfrag{#y8#}{$\smallhat_8$}
\psfrag{#y9#}{$\smallhat_9$}
\psfrag{#y10#}{$\smallhat_{10}$}
\psfrag{#y11#}{$\smallhat_{11}$}
\psfrag{#y12#}{$\smallhat_{12}$}
\psfrag{#z1#}{$z_{1}$}
\psfrag{#z2#}{$z_{2}$}
\psfrag{#z3#}{$z_{3}$}
\psfrag{#z4#}{$z_{4}$}
\psfrag{#z5#}{$z_{5}$}
\psfrag{#z6#}{$z_{6}$}
\psfrag{#z7#}{$z_{7}$}
\psfrag{#z8#}{$z_{8}$}
\psfrag{#z9#}{$z_{9}$}
\widgraph{.54\textwidth}{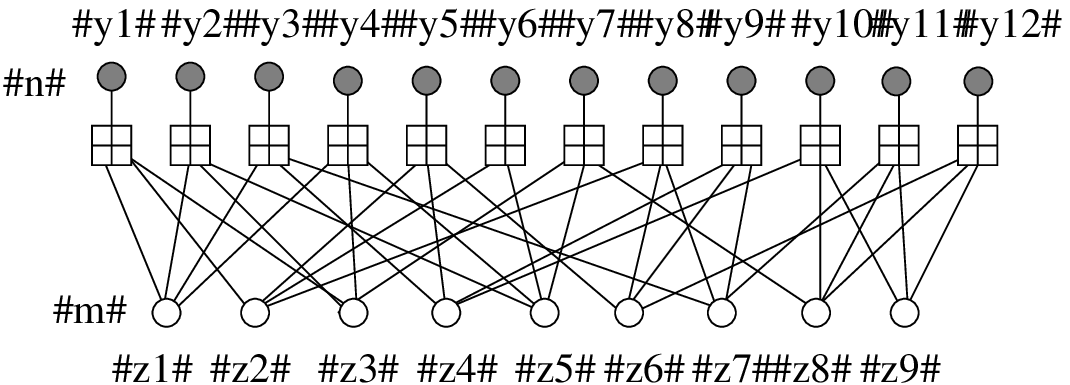} \\
(a) & (b)
\end{tabular}
\caption{(a) Factor graph representation of a rate $\rate = 0.5$
low-density parity check (LDPC) code with bit degree $\vdeg = 3$ and
check degree $\lowcdeg = 6$.  (b) Factor graph representation of a rate
$R = 0.75$ low-density generator matrix (LDGM) code with check degree
$\topdeg = 3$ and bit degree $\vdeg = 4$.}
\label{FigBasicCodes}
\end{center}
\end{figure}

Given a binary linear code $\mycode$, specified by parity check matrix
$\Parmat$, the code structure can be captured by a bipartite graph, in
which circular nodes ({\LARGE{$\circ$}}) represent the binary values
$x_i$ (or columns of $\Parmat$), and square nodes ($\blacksquare$)
represent the parity checks (or rows of $\Parmat$).  For instance,
Fig.~\ref{FigBasicCodes}(a) shows the factor graph for a rate $\rate =
\frac{1}{2}$ code in parity check form, with $\numcheck = 6$ checks
acting on $\numbit = 12$ bits.  The edges in this graph correspond to
$1$'s in the parity check matrix, and reveal the subset of bits on
which each parity check acts.  The parity check code in
Fig.~\ref{FigBasicCodes}(a) is a regular code with bit degree $3$ and
check degree $6$.  Such \emph{low-density} constructions, meaning that
both the bit degrees and check degrees remain bounded independently of
the block length $\numbit$, are of most practical use, since they can
be efficiently represented and stored, and yield excellent performance
under message-passing decoding.  In the context of a channel coding
problem, the shaded circular nodes at the top of the \emph{low-density
parity check} (LDPC) code in panel (a) represent the observed
variables $y_i$ received from the noisy channel.

Figure~\ref{FigBasicCodes}(b) shows a binary linear code represented
in factor graph form by its generator matrix $\Genmat$.  In this dual
representation, each codeword $x \in \{0,1\}^\nbit$ is generated by
taking the matrix-vector product of the form $\Genmat z$, where $z \in
\{0,1\}^m$ is a sequence of information bits, and $\Genmat \in
\{0,1\}^{\nbit \times m}$ is the generator matrix.  For the code shown
in panel (b), the blocklength is $n = 12$, and information sequences
are of length $m = 9$, for an overall rate of $\myrate = m/n = 0.75$
in this case.  The degrees of the check and variable nodes in the
factor graph are $\topdeg = 3$ and $\vdeg = 4$ respectively, so that
the associated generator matrix $\Genmat$ has $\topdeg = 3$ ones in
each row, and $\vdeg = 4$ ones in each column.  When the generator
matrix is sparse in this setting, then the resulting code is known as
a \emph{low-density generator matrix} (LDGM) code.

\subsection{Weight enumerating functions}

For future reference, it is useful to define the weight enumerating
function of a code.  Given a binary linear code of blocklength
$\numinfobit$, its codewords $x$ have renormalized Hamming weights
$\mywei \mydefn \frac{\|x\|_1}{\numinfobit}$ that range in the
interval $[0,1]$.  Accordingly, it is convenient to define a function
$\MyWtEnum_\numinfobit:[0,1] \rightarrow \real_+$ that, for each
$\mywei \in [0,1]$, counts the number of codewords of weight $\mywei$:
\begin{eqnarray}
\label{EqnDefnMyWtEnum}
\MyWtEnum_\numinfobit(\mywei) & \mydefn & \left| \left \{ x \in
\mycode \; \mid \; \mywei = \left \lceil \frac{\|x\|_1}{\numinfobit}
\right \rceil \; \right \} \right |,
\end{eqnarray}
where $\lceil \cdot \rceil$ denotes the ceiling function.  Although it
is typically difficult to compute the weight enumerator itself, it is
frequently possible to compute (or bound) the \emph{average weight
enumerator,} where the expectation is taken over some random ensemble
of codes.  In particular, our analysis in the sequel makes use of the
average weight enumerator of a $(\vdeg, \lowcdeg)$-regular LDPC code
(see~\figref{FigBasicCodes}(a)), defined as
\begin{eqnarray}
\label{EqnDefnAvWtEnum} 
\AvWtEnum{\numinfobit}(\mywei; \vdeg, \lowcdeg) & \mydefn &
\frac{1}{\numinfobit} \log \Exs \left[\MyWtEnum_\numinfobit(\mywei)
\right],
\end{eqnarray}
where the expectation is taken over the ensemble of all regular
$(\vdeg, \lowcdeg)$-LDPC codes.  For such regular LDPC codes, this
average weight enumerator has been extensively studied in previous
work~\cite{Gallager63,Litsyn02}.

\section{Optimality of bounded degree compound constructions}
\label{SecGenConstruc}

In this section, we describe the compound LDGM/LDPC construction that
is the focus of this paper, and describe our main results on their
source and channel coding optimality.

\subsection{Compound construction}

\begin{figure}[h]
\begin{center}
\psfrag{#k#}{$\kval$} \psfrag{#k1#}{$\kval_1$}
\psfrag{#k2#}{$\kval_2$} \psfrag{#n#}{$n$} \psfrag{#m#}{$m$}
\psfrag{#H#}{$\Parmat$} \psfrag{#H1#}{$\Parmat_1$}
\psfrag{#H2#}{$\Parmat_2$} \psfrag{#G#}{$\Genmat$}
\psfrag{#topdeg#}{$\cdeg$} \psfrag{#cdeg#}{$\lowcdeg$}
\psfrag{#vdeg#}{$\vdeg$} \raisebox{0.4in}{
\widgraph{0.6\textwidth}{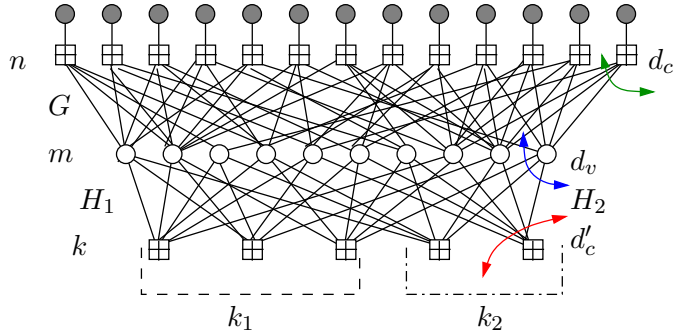}}
\end{center}
\caption{The compound LDGM/LDPC construction analyzed in this paper,
consisting of a $(\numbit, \numcheck)$ LDGM code over the middle and
top layers, compounded with a $(\numcheck, \kval)$ LDPC code over the
middle and bottom layers.  Codewords $x \in \{0,1\}^\numcodebit$ are
placed on the top row of the construction, and are associated with
information bit sequences $z \in \{0,1\}^\numinfobit$ in the middle
layer. The LDGM code over the top and middle layers is defined by a
sparse generator matrix $\Genmat \in \{0,1\}^{\numcodebit \times
\numinfobit}$ with at most $\topdeg$ ones per row.  The bottom LDPC
over the middle and bottom layers is represented by a sparse parity
check matrix $\Parmat \in \{0,1\}^{\kval \times \numinfobit}$ with
$\vdeg$ ones per column, and $\lowcdeg$ ones per row.}
\label{FigCompound}
\end{figure}

Our main focus is the construction illustrated
in~\figref{FigCompound}, obtained by compounding an LDGM code (top two
layers) with an LDPC code (bottom two layers).  The code is defined by
a factor graph with three layers: at the top, a vector $\codebit \in
\{0,1\}^\numcodebit$ of codeword bits is connected to a set of
$\numcodebit$ parity checks, which are in turn connected by a sparse
generator matrix $\Genmat$ to a vector $\infobit \in
\{0,1\}^\numinfobit$ of information bits in the middle layer.  The
information bits $\infobit$ are also codewords in an LDPC code,
defined by the parity check matrix $\Parmat$ connecting the middle and
bottom layers.

In more detail, considering first the LDGM component of the compound
code, each codeword $x \in \{0,1\}^\numcodebit$ in the top layer is
connected via the generator matrix $\Genmat \in \{0,1\}^{\numcodebit
\times \numinfobit}$ to an information sequence $y \in
\{0,1\}^\numinfobit$ in the middle layer; more specifically, we have
the algebraic relation $x = \Genmat y$.  Note that this LDGM code has
rate $\rateldgm \leq \frac{\numinfobit}{\numcodebit}$.  Second,
turning to the LDPC component of the compound construction, its
codewords correspond to a subset of information sequences $y \in
\{0,1\}^\numinfobit$ in the middle layer.  In particular, any valid
codeword $y$ satisfies the parity check relation $\Parmat y = 0$,
where $\Parmat \in \{0,1\}^{\numinfobit \times \kval}$ joins the
middle and bottom layers of the construction.  Overall, this defines
an LDPC code with rate $\rateldpc = 1 - \frac{\kval}{\numinfobit}$,
assuming that $\Parmat$ has full row rank.

The overall code $\mycode$ obtained by concatenating the LDGM and LDPC
codes has blocklength $\numcodebit$, and rate $\ratetot$ upper bounded
by $\rateldgm \rateldpc$.  In algebraic terms, the code $\mycode$ is
defined as
\begin{eqnarray}
\mycode & \mydefn & \left \{ \codebit \in \{0,1\}^\numcodebit \; \mid \;
\codebit = \Genmat \infobit \quad \mbox{for some $\infobit \in
\{0,1\}^\numinfobit$ such that} \quad \Parmat \infobit = 0 \right \},
\end{eqnarray}
where all operations are in modulo two arithmetic.

Our analysis in this paper will be performed over random ensembles of
compound LDGM/LDPC ensembles.  In particular, for each degree triplet
$(\topdeg, \vdeg, \lowcdeg)$, we focus on the following random
ensemble: 
\begin{enumerate}
\item[(a)] For each fixed integer $\topdeg \geq 4$, the random
generator matrix $\Genmat \in \{0,1\}^{\numcodebit \times
\numinfobit}$ is specified as follows: for each of the $\numcodebit$
rows, we choose $\topdeg$ positions with replacement, and put a $1$ in
each of these positions.  This procedure yields a random matrix with
at most $\topdeg$ ones per row, since it is possible (although of
asymptotically negligible probability for any fixed $\topdeg$) that
the same position is chosen more than once.
\item[(b)] For each fixed degree pair $(\vdeg, \lowcdeg)$, the random
LDPC matrix $\Parmat \in \{0,1\}^{\kval \times \numinfobit}$ is chosen
uniformly at random from the space of all matrices with exactly
$\vdeg$ ones per column, and exactly $\lowcdeg$ ones per row.  This
ensemble is a standard $(\vdeg, \lowcdeg)$-regular LDPC ensemble.
\end{enumerate}
We note that our reason for choosing the check-regular LDGM ensemble
specified in step (a) is not that it need define a particularly good
code, but rather that it is convenient for theoretical purposes.
Interestingly, our analysis shows that the bounded degree $\topdeg$
check-regular LDGM ensemble, even though it is sub-optimal for both
source and channel coding in isolation~\cite{MarWai06a,MarWai06b},
defines optimal source and channel codes when combined with a bottom
LDPC code.

\subsection{Main results}

Our first main result is on the achievability of the Shannon
rate-distortion bound using codes from LDGM/LDPC compound construction
with \emph{finite degrees} $(\cdeg, \vdeg, \lowcdeg)$.  In particular,
we make the following claim:
\btheos
\label{ThmSource}
Given any pair $(\myrate, \distor)$ satisfying the Shannon bound,
there is a set of finite degrees $(\cdeg, \vdeg, \lowcdeg)$ and a code
from the associated LDGM/LDPC ensemble with rate $\myrate$ that is a
$\distor$-good source code (see Definition~\ref{DefnCoding}).
\etheos
In other work~\cite{MarWai06a,MarWai06c}, we showed that standard LDGM
codes from the check-regular ensemble cannot achieve the
rate-distortion bound with finite degrees.  As will be highlighted by
the proof of Theorem~\ref{ThmSource} in
Section~\ref{SecSourceOptimal}, the inclusion of the LDPC lower code
in the compound construction plays a vital role in the achievability
of the Shannon rate-distortion curve.

Our second main result of this result is complementary in nature to
Theorem~\ref{ThmSource}, regarding the achievability of the Shannon
channel capacity using codes from LDGM/LDPC compound construction with
\emph{finite degrees} $(\cdeg, \vdeg, \lowcdeg)$.  In particular, we
have:
\btheos
\label{ThmChannel}
For all rate-noise pairs $(\myrate, \channoise)$ satisfying the
Shannon channel coding bound $\myrate < 1 - \binent{\channoise}$,
there is a set of finite degrees $(\cdeg, \vdeg, \lowcdeg)$ and a code
from the associated LDGM/LDPC ensemble with rate $\myrate$ that is a
$\channoise$-good channel code (see Definition~\ref{DefnCoding}).
\etheos
\noindent To put this result into perspective, recall that the overall
rate of this compound construction is given by $\ratetot = \rateldgm
\rateldpc$.  Note that an LDGM code on its own (i.e., without the
lower LDPC code) is a special case of this construction with
$\rateldpc = 1$.  However, a standard LDGM of this variety is
\emph{not} a good channel code, due to the large number of low-weight
codewords.  Essentially, the proof of Theorem~\ref{ThmChannel} (see
Section~\ref{SecChannelOptimal}) shows that using a non-trivial LDPC
lower code (with $\rateldpc < 1$) can eliminate these troublesome
low-weight codewords.


\section{Consequences for coding with side information}
\label{SecSideInformation}

We now turn to consideration of the consequences of our two main
results for problems of coding with side information.  It is
well-known from previous work~\cite{Zamir02} that achieving the
information-theoretic limits for these problems requires nested
constructions, in which a collection of good source codes are nested
inside a good channel code (or vice versa).  Accordingly, we begin in
Section~\ref{SecNesting} by describing how our compound construction
naturally generates such nested ensembles.  In Sections~\ref{SecSCSI}
and~\ref{SecCCSI} respectively, we discuss how the compound
construction can be used to achieve the information-theoretic optimum
for binary source coding with side information (a version of the
Wyner-Ziv problem~\cite{WynerZiv76}), and binary information embedding
(a version of ``dirty paper coding'', or the Gelfand-Pinsker
problem~\cite{Gelfand80}).

\subsection{Nested code structure}
\label{SecNesting}

The structure of the compound LDGM/LDPC construction lends itself
naturally to nested code constructions.  In particular, we first
partition the set of $\kval$ lower parity checks into two disjoint
subsets $\Kset_1$ and $\Kset_2$, of sizes $\kval_1$ and $\kval_2$
respectively, as illustrated in~\figref{FigCompound}.  Let $\Parmat_1$
and $\Parmat_2$ denote the corresponding partitions of the full parity
check matrix $\Parmat \in \{0,1\}^{\kval \times \midbit}$.  Now let us
set all parity bits in the subset $\Kset_2$ equal to zero, and
consider the LDGM/LDPC code $\mycodepar$ defined by the generator
matrix $\Genmat$ and the parity check (sub)matrix $\Parmat_1$, as
follows
\begin{eqnarray}
\label{EqnDefnMycodepar}
\mycodepar & \mydefn & \left \{ \codebit \in \{0,1\}^\numcodebit \;
\mid \; \codebit = \Genmat \infobit \quad \mbox{for some $\infobit \in
\{0,1\}^\numinfobit$ such that} \quad \Parmat_1 \; \infobit = 0 \right
\}.
\end{eqnarray}
Note that the rate of $\mycodepar$ is given by $\rate' = \rateldgm \,
\rateldpcone$, which can be suitably adjusted by modifying the LDGM
and LDPC rates respectively.  Moreover, by applying
Theorems~\ref{ThmSource} and~\ref{ThmChannel}, there exist finite
choices of degree such that $\mycodepar$ will be optimal for both
source and channel coding.

Considering now the remaining $\kval_2$ parity bits in the subset
$\Kset_2$, suppose that we set them equal to a fixed binary sequence
$\msca \in \{0,1\}^{\kval_2}$.  Now consider the code
\begin{eqnarray}
\mycode(\msca) & \mydefn & \left \{ \codebit \in \{0,1\}^\numcodebit
\; \mid \; \codebit = \Genmat \infobit \quad \mbox{for some $\infobit
\in \{0,1\}^\numinfobit$ such that} \quad \begin{bmatrix} \Parmat_1 \\
\Parmat_2 \end{bmatrix} \; \infobit = \begin{bmatrix} 0 \\ \msca
\end{bmatrix} \; \right \}.
\end{eqnarray}
Note that for each binary sequence $\msca \in \{0,1\}^{\kval_2}$, the
code $\mycode(\msca)$ is a subcode of $\mycodepar$; moreover, the
collection of these subcodes forms a disjoint partition as follows
\begin{eqnarray}
\label{EqnDisPartition}
\mycodepar & = & \bigcup_{\msca \in \{0,1\}^{\kval_2}} \mycode(\msca).
\end{eqnarray}
Again, Theorems~\ref{ThmSource} and~\ref{ThmChannel} guarantee that
(with suitable degree choices), each of the subcodes $\mycode(\msca)$
is optimal for both source and channel coding.  Thus, the LDGM/LDPC
construction has a natural nested property, in which a good
source/channel code---namely $\mycodepar$---is partitioned into a
disjoint collection $\{ \mycode(\msca), \; \msca \in \{0,1\}^{\kval_1}
\}$ of good source/channel codes.  We now illustrate how this nested
structure can be exploited for coding with side information.

\subsection{Source coding with side information}
\label{SecSCSI}

We begin by showing that the compound construction can be used
to perform source coding with side information (SCSI).

\subsubsection{Problem formulation}

\begin{figure}[h]
\begin{center}
\psfrag{#s#}{$\Ssca$} \psfrag{#shat#}{$\estim{\Ssca}$}
\psfrag{#w#}{$\Wnoise$} \psfrag{#y#}{$\Wzside$}
\psfrag{#rate#}{$\myrate$}
\widgraph{0.5\textwidth}{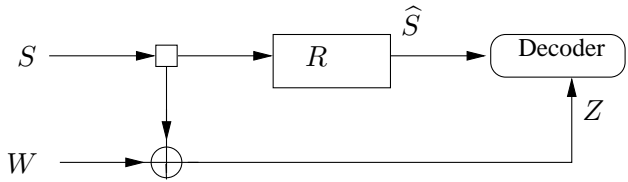}
\caption{Block diagram representation of source coding with side
information (SCSI).  A source $\Ssca$ is compressed to rate $\myrate$.  The
decoder is given the compressed version, and side information $\Wzside
= \Ssca \oplus \Wnoise$, and wishes to use $(\estim{\Ssca}, \Wzside)$
to reconstruct the source $\Ssca$ up to distortion $\distor$.  }
\label{FigWynerZiv}
\end{center}
\end{figure}

Suppose that we wish to compress a symmetric Bernoulli source $\Ssca
\sim \myber(\frac{1}{2})$ so as to be able to reconstruct it with
Hamming distortion $\distor$.  As discussed earlier in
Section~\ref{SecBackground}, the minimum achievable rate is given by
$\myrate(\distor) = 1 - \binent{\distor}$.  In the Wyner-Ziv extension
of standard lossy compression~\cite{WynerZiv76}, there is an
additional source of side information about $\Ssca$---say in the form
\mbox{$\Wzside = \Ssca \oplus \Wnoise$} where $\Wnoise \sim
\myber(\bernoise)$ is observation noise---that is available only at
the decoder.  See~\figref{FigWynerZiv} for a block diagram
representation of this problem.

For this binary version of source coding with side information (SCSI),
it is known~\cite{Barron03} that the minimum achievable rate takes the
form
\begin{eqnarray}
\label{EqnDefnWynerZiv}
\ratewz(\distor, \channoise) = \lce \big \{ \binent{\distor \ast
\channoise} - \binent{\distor}, \, (\channoise,0) \big \},
\end{eqnarray}
where $\lce$ denotes the lower convex envelope.  Note that in the
special case $\channoise = \frac{1}{2}$, the side information is
useless, so that the Wyner-Ziv rate reduces to classical
rate-distortion.  In the discussion to follow, we focus only on
achieving rates of the form $\binent{\distor \ast \channoise} -
\binent{\distor}$, as any remaining rates on the Wyner-Ziv
curve~\eqref{EqnDefnWynerZiv} can be achieved by time-sharing with the
point $(\channoise, 0)$.

\subsubsection{Coding procedure for SCSI}

In order to achieve rates of the form $\myrate = \binent{\distor \ast
\channoise} - \binent{\distor}$, we use the compound LDGM/LDPC
construction, as illustrated in~\figref{FigCompound}, according to the
following procedure. \\

\begin{figure}[h]
\begin{center}
\begin{tabular}{cc}
\psfrag{#b1#}{$0$}
\psfrag{#b2#}{$0$}
\psfrag{#b3#}{$0$}
\psfrag{#k1#}{$\kval_1$}
\psfrag{#n#}{$n$} \psfrag{#m#}{$m$}
\psfrag{#H1#}{$\Parmat_1$}
\psfrag{#G#}{$\Genmat$}
\psfrag{#topdeg#}{$\cdeg$} \psfrag{#cdeg#}{$\lowcdeg$}
\psfrag{#vdeg#}{$\vdeg$} 
\widgraph{.5\textwidth}{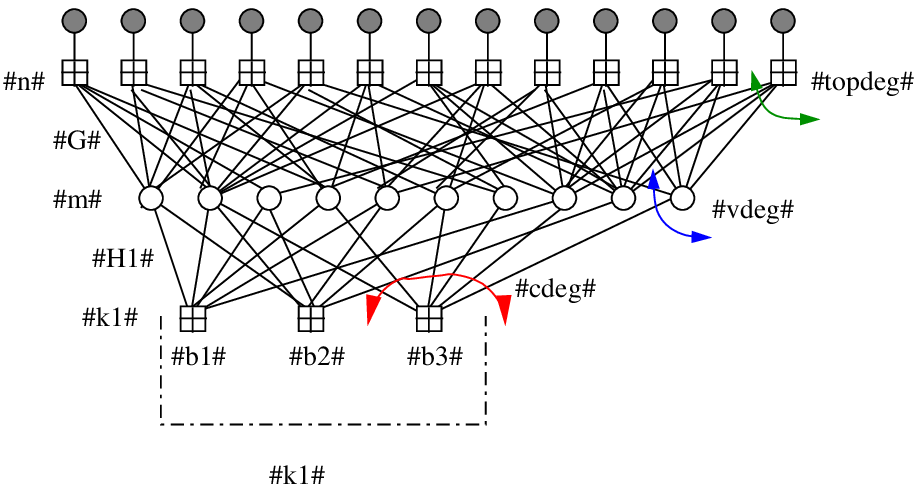} &
\psfrag{#b1#}{$0$}
\psfrag{#b2#}{$0$}
\psfrag{#b3#}{$0$}
\psfrag{#b4#}{$1$}
\psfrag{#b5#}{$0$}
\psfrag{#k1#}{$\kval_1$}
\psfrag{#k2#}{$\kval_2$}
\psfrag{#k#}{$\kval$}
\psfrag{#n#}{$n$} \psfrag{#m#}{$m$}
\psfrag{#H1#}{$\Parmat_1$}
\psfrag{#H2#}{$\Parmat_2$}
\psfrag{#G#}{$\Genmat$}
\psfrag{#topdeg#}{$\cdeg$} \psfrag{#cdeg#}{$\lowcdeg$}
\psfrag{#vdeg#}{$\vdeg$} 
\widgraph{.5\textwidth}{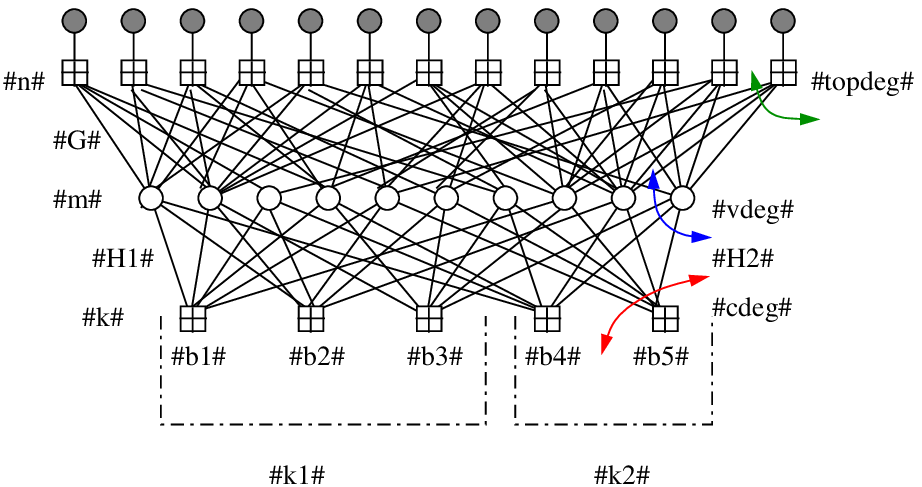} \\
(a) & (b)
\end{tabular}
\caption{(a) Source coding stage for Wyner-Ziv procedure: the
$\mycodepar$, specified by the generator matrix $\Genmat \in
\{0,1\}^{\topbit \times \numinfobit}$ and parity check matrix
$\Parmat_1 \in \{0,1\}^{\kval_1 \times \numinfobit}$, is used to
quantize the source vector $\Ssca \in \{0,1\}^\topbit$, thereby
obtaining a quantized version $\estim{\Ssca} \in \{0,1\}^\topbit$ and
associated vector of information bits $\estim{\Ysca} \in
\{0,1\}^\numinfobit$, such that $\estim{\Ssca} = \Genmat \,
\estim{\Ysca}$ and $\Parmat_1 \, \estim{\Ysca} = 0$.}
\label{FigWZCoding}
\end{center}
\end{figure}

\mypara{Step \#1, Source coding:} The first step is a source coding
operation, in which we transform the source sequence $\Ssca$ to a
quantized representation $\Ssca$.  In order to do so, we use the code
$\mycodepar$, as defined in equation~\eqref{EqnDefnMycodepar} and
illustrated in~\figref{FigWZCoding}(a), composed of the generator
matrix $\Genmat$ and the parity check matrix $\Parmat_1$.  Note that
$\mycodepar$, when viewed as a code with blocklength $\topbit$, has
rate $\rate_1 \mydefn \frac{\midbit \, \big(1 -
\frac{\lowbit_1}{\midbit}\big)}{\nbit} \; = \;
\frac{\midbit-\lowbit_1}{\nbit}$.  Suppose that we
choose\footnote{Note that the choices of $\midbit$ and $\lowbit_1$
need not be unique.} the middle and lower layer sizes $\midbit$ and
$\lowbit_1$ respectively such that
\begin{eqnarray}
\label{EqnDefnRateoneWZ}
\rate_1 & = & \frac{\midbit-\lowbit_1}{\nbit} \; = \; 1 -
\binent{\distor} + \epsilon/2,
\end{eqnarray}
where $\epsilon > 0$ is arbitrary.  For any such choice,
Theorem~\ref{ThmSource} guarantees the existence of finite degrees
$(\topdeg, \vdeg, \lowcdeg)$ such that that $\mycodepar$ is a good
$\distor$-distortion source code.  Consequently, for the specified
rate $\rate_1$, we can use $\mycodepar$ in order to transform the
source to some quantized representation $\estim{\Ssca}$ such that the
error $\estim{\Ssca} \oplus \Ssca$ has average Hamming weighted
bounded by $\distor$. Moreover, since $\estim{\Ssca}$ is a codeword of
$\mycodepar$, there is some sequence of information bits
$\estim{\Ysca} \in \{0,1\}^\numinfobit$ such that $\estim{\Ssca} =
\Genmat \estim{\Ysca}$ and $\Parmat_1 \estim{\Ysca} = 0$. \\

\mypara{Step \#2.  Channel coding:} Given the output $(\estim{\Ysca},
\estim{\Ssca})$ of the source coding step, consider the sequence
$\Parmat_2 \estim{\Ysca} \in \{0,1\}^{\lowbit_2}$ of parity bits
associated with the parity check matrix $\Parmat_2$.  Transmitting
this string of parity bits requires rate $\rateeff =
\frac{\lowbit_2}{\topbit}$.  Overall, the decoder receives both these
$\lowbit_2$ parity bits, as well as the side information sequence
$\Wzside = \Ssca \oplus \Wnoise$.  Using these two pieces of
information, the goal of the decoder is to recover the quantized
sequence $\estim{\Ssca}$.  

Viewing this problem as one of channel coding, the effective rate of
this channel code is \mbox{$\rate_2 = \frac{\midbit - \lowbit_1 -
\lowbit_2}{\topbit}$.}  Note that the side information can be written
in the form
\begin{eqnarray*}
\Wzside & = & \Ssca \oplus \Wnoise \; = \; \estim{\Ssca} \oplus
\Qnoise \oplus \Wnoise,
\end{eqnarray*}
where $\Qnoise \mydefn \Ssca \oplus \estim{\Ssca}$ is the quantization
noise, and $\Wnoise \sim \myber(\channoise)$ is the channel noise.  If
the quantization noise $\Qnoise$ were i.i.d. $\myber(\distor)$, then
the overall effective noise $\Qnoise\oplus \Wnoise$ would be
i.i.d. $\myber(\distor \ast \channoise)$.  (In reality, the
quantization noise is not exactly i.i.d. $\myber(\distor)$, but it can
be shown~\cite{Zamir02} that it can be treated as such for theoretical
purposes.)  Consequently, if we choose $\lowbit_2$ such that
\begin{eqnarray}
\label{EqnDefnRatetwoWZ}
\rate_2 & = & \frac{\midbit - \lowbit_1 - \lowbit_2}{\topbit} \; = \;
1 - \binent{\distor \ast \channoise} - \epsilon/2,
\end{eqnarray}
for an arbitrary $\epsilon > 0$, then Theorem~\ref{ThmChannel}
guarantees that the decoder will (w.h.p.) be able to recover a
codeword corrupted by $(\distor \ast \channoise)$-Bernoulli noise.  \\

\vspace*{.02in}

\noindent Summarizing our findings, we state the following:
\bcors
There exist finite choices of degrees $(\topdeg, \vdeg, \lowcdeg)$
such that the compound LDGM/LDPC construction achieves the Wyner-Ziv
bound.
\ecors
\begin{proof}
With the source coding rate $\rate_1$ chosen according to
equation~\eqref{EqnDefnRateoneWZ}, the encoder will return a
quantization $\estim{\Ssca}$ with average Hamming distance to the
source $\Ssca$ of at most $\distor$.  With the channel coding rate
$\rate_2$ chosen according to equation~\eqref{EqnDefnRatetwoWZ}, the
decoder can with high probability recover the quantization
$\estim{\Ssca}$.  The overall transmission rate of the scheme is
\begin{eqnarray*}
\rateeff & = & \frac{\lowbit_2}{\topbit} \\
& = & \frac{\midbit-\lowbit_1}{\nbit} - \frac{\midbit - \lowbit_1 -
\lowbit_2}{\topbit} \\ 
& = & \rate_1 - \rate_2 \\
& = & \left(1 - \binent{\distor} + \epsilon/2 \right) - \left(1 -
\binent{\distor \ast \channoise} - \epsilon/2 \right) \\ 
& = & \binent{\distor \ast \channoise} - \binent{\distor} + \epsilon.
\end{eqnarray*}
Since $\epsilon > 0$ was arbitrary, we have established that the
scheme can achieve rates arbitrarily close to the Wyner-Ziv bound.
\end{proof}
%


\subsection{Channel coding with side information}
\label{SecCCSI}

We now show how the compound construction can be used to perform
channel coding with side information (CCSI).

\subsubsection{Problem formulation}

In the binary information embedding problem, given a specified input
vector $\Chaninput \in \{0,1\}^\topbit$, the channel output
$\Chanoutput \in \{0,1\}^\topbit$ is assumed to take the form
\begin{eqnarray}
\label{EqnGPChannel}
\Chanoutput & = & \Chaninput \oplus \Shost \oplus \Wnoise,
\end{eqnarray}
where $\Shost$ is a host signal (not under control of the user), and
$\Wnoise \sim \myber(\channoise)$ corresponds to channel noise.  The
encoder is free to choose the input vector $\Chaninput \in
\{0,1\}^\topbit$, subject to an average channel
constraint 
\begin{eqnarray}
\label{EqnGPConstraint}
\frac{1}{\topbit} \Exs \left[ \|\Chaninput \|_1 \right] & \leq & \iew,
\end{eqnarray}
for some parameter $\iew \in (0, \myhalf]$.  The goal is to use a
channel coding scheme that satisfies this
constraint~\eqref{EqnGPConstraint} so as to maximize the number of
possible messages $\Gpmess$ that can be reliably communicated.
Moreover, We write $\Chaninput \equiv \Chaninput_{\Gpmess}$ to
indicate that each channel input is implicitly identified with some
underlying message $\Gpmess$.  Given the channel output $\Chanoutput =
\Chaninput_{\Gpmess} \oplus \Shost \oplus \Wnoise$, the goal of the
decoder is to recover the embedded message $\Gpmess$.  The capacity
for this binary information embedding problem~\cite{Barron03} is given
by
\begin{eqnarray}
\label{EqnGPRate}
\rateie(\iew, \channoise) & = & \uce \big\{ \binent{\iew} -
\binent{\channoise}, (0,0) \big \},
\end{eqnarray}
where $\uce$ denotes the upper convex envelope.  As before, we focus
on achieving rates of the form $\binent{\iew} - \binent{\channoise}$,
since any remaining points on the curve~\eqref{EqnGPRate} can be
achieved via time-sharing with the $(0,0)$ point.

\begin{figure}[h]
\begin{center}
\psfrag{#m#}{$\gpmsca$}
\psfrag{#mhat#}{$\estim{\gpmsca}$}
\psfrag{#s#}{$\Shost$} 
\psfrag{#u#}{$\Chaninput_\gpmsca$}
\psfrag{#w#}{$\Wnoise$} 
\psfrag{#co#}{$\Chanoutput$}
\psfrag{#iew#}{}
\widgraph{0.75\textwidth}{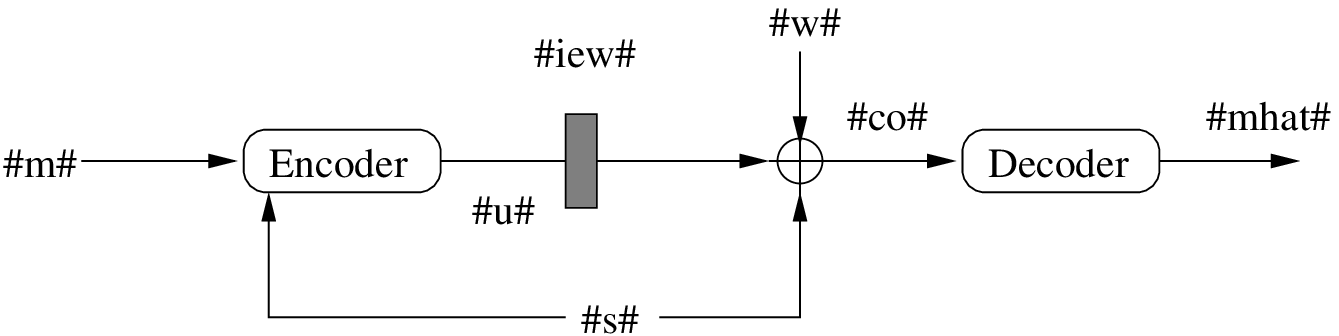}
\caption{Block diagram representation of channel coding with side
information (CCSI).  The encoder embeds a message $\gpmsca$ into the
channel input $\Chaninput_\msca$, which is required to satisfy the
average channel constraint $\frac{1}{\topbit}
\Exs[\|\Chaninput_\gpmsca\|_1] \leq \iew$.  The channel produces the
output $\Chanoutput = \Chaninput_\gpmsca \oplus \Shost \oplus
\Wnoise$, where $\Shost$ is a host signal known only to the encoder,
and $\Wnoise \sim \myber(\channoise)$ is channel noise.  Given the
channel output $\Ysca$, the decoder outputs an estimate
$\estim{\gpmsca}$ of the embedded message.}
\label{FigInfoEmbed}
\end{center}
\end{figure}

\subsubsection{Coding procedure for CCSI}

In order to achieve rates of the form $\myrate = \binent{\iew} -
\binent{\channoise}$, we again use the compound LDGM/LDPC construction
in~\figref{FigCompound}, now according to the following two step
procedure. \\

\mypara{Step \#1: Source coding:} The goal of the first stage is to
embed the message into the transmitted signal $\Chaninput$ via a
quantization process.  In order to do so, we use the code illustrated
in~\figref{FigGPCoding}(a), specified by the generator matrix
$\Genmat$ and parity check matrices $\Parmat_1$ and $\Parmat_2$.  The
set $\Kset_1$ of $\lowbit_1$ parity bits associated with the check
matrix $\Parmat_1$ remain fixed to zero throughout the scheme.  On the
other hand, we use the remaining $\lowbit_2$ lower parity bits
associated with $\Parmat_2$ to specify a particular message $\gpmsca \in
\{0,1\}^{\lowbit_2}$ that the decoder would like to recover.  In algebraic
terms, the resulting code $\mycodegp(\gpmsca)$ has the form
\begin{eqnarray}
\label{EqnDefnGPsourcecode}
\mycodegp(\gpmsca) & \mydefn & \left \{ \codebit \in \{0,1\}^\numcodebit
\; \mid \; \codebit = \Genmat \infobit \quad \mbox{for some $\infobit
\in \{0,1\}^\numinfobit$ such that} \quad \begin{bmatrix} \Parmat_1 \\
\Parmat_2 \end{bmatrix} \; \infobit = \begin{bmatrix} 0 \\ \gpmsca
\end{bmatrix} \; \right \}.
\end{eqnarray}
Since the encoder has access to host signal $\Ssca$, it may use this
code $\mycodegp(\gpmsca)$ in order to quantize the host signal.  After
doing so, the encoder has a quantized signal $\estim{\Ssca}_\gpmsca \in
\{0,1\}^\topbit$ and an associated sequence $\estim{\Ysca}_\gpmsca \in
\{0,1\}^\numinfobit$ of information bits such that
$\estim{\Ssca}_\gpmsca = \Genmat \, \estim{\Ysca}_\gpmsca$.  Note that the
quantized signal $(\estim{\Ysca}_\gpmsca, \estim{\Ssca}_\gpmsca)$
specifies the message $\gpmsca$ in an implicit manner, since $\gpmsca =
\Parmat_2 \, \estim{\Ysca}_\gpmsca$ by construction of the code
$\mycodegp(\gpmsca)$.

Now suppose that we choose $\topbit, \midbit$ and $\lowbit$ such that
\begin{eqnarray}
\label{EqnGPSourceRate}
\rate_1 & = & \frac{\midbit - \lowbit_1 - \lowbit_2}{\topbit} \; = \;
1 - \binent{\iew} + \epsilon/2
\end{eqnarray}
for some $\epsilon > 0$, then Theorem~\ref{ThmSource} guarantees that
there exist finite degrees $(\topdeg, \vdeg, \lowcdeg)$ such that the
resulting code is a good $\iew$-distortion source code.  Otherwise
stated, we are guaranteed that w.h.p, the quantization error $\Qnoise
\mydefn \Ssca \oplus \estim{\Ssca}$ has average Hamming weight upper
bounded by $\iew \topbit$.  Consequently, we may set the channel input
$\Chaninput$ equal to the quantization noise ($\Chaninput = \Qnoise$),
thereby ensuring that the average channel
constraint~\eqref{EqnGPConstraint} is satisfied. \\

\begin{figure}[h]
\begin{center}
\begin{tabular}{cc}
\psfrag{#b1#}{$0$}
\psfrag{#b2#}{$0$}
\psfrag{#b3#}{$0$}
\psfrag{#b4#}{$1$}
\psfrag{#b5#}{$0$}
\psfrag{#k1#}{}
\psfrag{#k2#}{$\gpmsca$}
\psfrag{#k#}{$\kval$}
\psfrag{#n#}{$n$} \psfrag{#m#}{$m$}
\psfrag{#H1#}{$\Parmat_1$}
\psfrag{#H2#}{$\Parmat_2$}
\psfrag{#G#}{$\Genmat$}
\psfrag{#topdeg#}{$\cdeg$} \psfrag{#cdeg#}{$\lowcdeg$}
\psfrag{#vdeg#}{$\vdeg$} 
\widgraph{.5\textwidth}{wzcode_two.eps} &
\psfrag{#b1#}{$0$}
\psfrag{#b2#}{$0$}
\psfrag{#b3#}{$0$}
\psfrag{#k1#}{$$}
\psfrag{#n#}{$n$} \psfrag{#m#}{$m$}
\psfrag{#H1#}{$\Parmat_1$}
\psfrag{#G#}{$\Genmat$}
\psfrag{#topdeg#}{$\cdeg$} \psfrag{#cdeg#}{$\lowcdeg$}
\psfrag{#vdeg#}{$\vdeg$} 
\widgraph{.5\textwidth}{wzcode_one.eps} \\
%
(a) & (b)
\end{tabular}
\caption{(a) Source coding step for binary information embedding.  The
message $\gpmsca \in \{0,1\}^{\kval_2}$ specifies a particular coset;
using this particular source code, the host signal $\Shost$ is
compressed to $\estim{\Shost}$, and the quantization error $\Qnoise =
\Shost \oplus \estim{\Shost}$ is transmitted over the constrained
channel.  (b) Channel coding step for binary information embedding.
The decoder receives $\Chanoutput = \estim{\Shost} \oplus \Wnoise$
where $\Wnoise \sim \myber(\channoise)$ is channel noise, and seeks to
recover $\estim{\Shost}$, and hence the embedded message $\gpmsca$
specifying the coset.  }
\label{FigGPCoding}
\end{center}
\end{figure}

\mypara{Step \#2, Channel coding:} In the second phase, the decoder is
given a noisy channel observation of the form 
\begin{eqnarray}
\Chanoutput & = & \Qnoise \oplus \Shost \oplus \Wnoise \; = \;
\estim{\Ssca} \oplus \Wnoise,
\end{eqnarray}
and its task is to recover $\estim{\Ssca}$.  In terms of the code
architecture, the $\lowbit_1$ lower parity bits remain set to zero;
the remaining $\lowbit_2$ parity bits, which represent the message
$\gpmsca$, are unknown to the coder.  The resulting code, as
illustrated illustrated in~\figref{FigGPCoding}(b), can be viewed as
channel code with effective rate $\frac{\midbit -
\lowbit_1}{\topbit}$.  Now suppose that we choose $\lowbit_1$ such
that the effective code used by the decoder has rate
\begin{eqnarray} 
\label{EqnChanRateGelfand}
\rate_2 & = & \frac{\midbit - \lowbit_1}{\topbit} = 1 -
\binent{\channoise} - \epsilon/2,
\end{eqnarray}
for some $\epsilon > 0$.  Since the channel noise $\Wnoise$ is
$\myber(\channoise)$ and the rate $\rate_2$ chosen according
to~\eqref{EqnChanRateGelfand}, Theorem~\ref{ThmChannel} guarantees
that the decoder will w.h.p. be able to recover the pair
$\estim{\Ssca}$ and $\estim{\Ysca}$.  Moreover, by design of the
quantization procedure, we have the equivalence $\gpmsca = \Parmat_2
\, \estim{\Ysca}$ so that a simple syndrome-forming procedure allows
the decoder to recover the hidden message.  

\vspace*{.02in}

Summarizing our findings, we state the following:
\bcors
There exist finite choices of degrees $(\topdeg, \vdeg, \lowcdeg)$
such that the compound LDGM/LDPC construction achieves the binary
information embedding (Gelfand-Pinsker) bound.
\ecors
\begin{proof}
With the source coding rate $\rate_1$ chosen according to
equation~\eqref{EqnGPSourceRate}, the encoder will return a
quantization $\estim{\Shost}$ of the host signal $\Shost$ with average
Hamming distortion upper bounded by $\iew$.  Consequently,
transmitting the quantization error $\Qnoise = \Shost \oplus
\estim{\Shost}$ will satisfy the average channel
constraint~\eqref{EqnGPConstraint}.  With the channel coding rate
$\rate_2$ chosen according to equation~\eqref{EqnChanRateGelfand}, the
decoder can with high probability recover the quantized signal
$\estim{\Shost}$, and hence the message $\gpmsca$.  Overall, the
scheme allows a total of $2^{\lowbit_2}$ distinct messages to be
embedded, so that the effective information embedding rate is
\begin{eqnarray*}
\rateeff & = & \frac{\lowbit_2}{\topbit} \\
& = & \frac{\midbit-\lowbit_1}{\nbit} - \frac{\midbit - \lowbit_1 -
\lowbit_2}{\topbit} \\ 
& = & \rate_2 - \rate_1 \\
& = & \left(1 - \binent{\channoise} - \epsilon/2 \right) - \left(1 -
\binent{\iew} + \epsilon/2 \right) \\
& = & \binent{\iew} - \binent{\channoise} + \epsilon,
\end{eqnarray*}
for some $\epsilon > 0$.  Thus, we have shown that the proposed scheme
achieves the binary information embedding bound~\eqref{EqnGPRate}.
\end{proof}


\section{Proof of source coding optimality}
\label{SecSourceOptimal}

This section is devoted to the proof of the previously stated
Theorem~\ref{ThmSource} on the source coding optimality of the
compound construction.

\subsection{Set-up}

In establishing a rate-distortion result such as
Theorem~\ref{ThmSource}, perhaps the most natural focus is the random
variable
\begin{eqnarray}
\label{EqnDefnRvdistcom}
\rvdistcom & \mydefn & \frac{1}{\numbit} \min_{x \in \mycode} \| x -
\Ssca \|_1,
\end{eqnarray}
corresponding to the (renormalized) minimum Hamming distance from a
random source sequence $\Ssca \in \{0,1 \}^\topbit$ to the nearest
codeword in the code $\mycode$.  Rather than analyzing this random
variable directly, our proof of Theorem~\ref{ThmSource} proceeds
indirectly, by studying an alternative random variable.

Given a binary linear code with $\numcw$ codewords, let $i = 0, 1, 2,
\ldots, \numcw-1$ be indices for the different codewords.  We say that
a codeword $\cw{i}$ is \emph{distortion $\distor$-good} for a source
sequence $\Ssca$ if the Hamming distance $\| \cw{i} \oplus \Ssca \|_1$
is at most $\distor \numbit$.  We then set the indicator random
variable $\cwind{i}(\distor) = 1$ when codeword $\cw{i}$ is distortion
$\distor$-good.  With these definitions, our proof is based on the
following random variable:
\begin{eqnarray}
\Complexnumc & \mydefn & \sum_{i=0}^{\numcw -1} \cwind{i}(\distor).
\end{eqnarray}
Note that $\Complexnumc$ simply counts the number of codewords that
are distortion $\distor$-good for a source sequence $\Ssca$.
Moreover, for all distortions $\distor$, the random variable
$\Complexnumc$ is linked to $\rvdistcom$ via the equivalence
\begin{eqnarray}
\label{EqnKeyEquiv}
\mprob[\Complexnumc > 0] & = & \mprob[\rvdistcom \leq \distor].
\end{eqnarray}

Throughout our analysis of $\mprob[\Complexnumc > 0]$, we carefully
track only its exponential behavior.  More precisely, the analysis to
follow will establish an inverse polynomial lower bound of the form
$\mprob[\Complexnumc > 0] \geq 1/f(\topbit)$ where $f(\cdot)$ collects
various polynomial factors.  The following concentration result
establishes that the polynomial factors in these bounds can be ignored:
\blems[Sharp concentration] Suppose that for some target distortion
$\distor$, we have
\begin{eqnarray}
\label{EqnInvPolyDecay}
\mprob[\Complexnumc > 0] & \geq &  1/f(\topbit),
\end{eqnarray}
where $f(\cdot)$ is a polynomial function satisfying $\log f(n) =
o(n)$.  Then for all $\epsilon > 0$, there exists a \emph{fixed code}
$\bar{\mycode}$ of sufficiently large blocklength $\topbit$ such that
$\Exs[d_n(\Ssca; \bar{\mycode})] \leq \distor + \epsilon$.
\elems
\begin{proof}
Let us denote the random code $\mycode$ as $(\mycodetop, \mycodebot)$,
where $\mycodetop$ denotes the random LDGM top code, and $\mycodebot$
denotes the random LDPC bottom code.  Throughout the analysis, we
condition on some fixed LDPC bottom code, say $\mycodebot =
\mycodebarbot$.  We begin by showing that the random variable
$(\rvdistcom \; | \; \mycodebarbot)$ is sharply concentrated.  In order
to do so, we construct a vertex-exposure martingale~\cite{Motwani95}
of the following form.  Consider a fixed sequential labelling $\{1,
\ldots, \numbit \}$ of the top LDGM checks, with check $i$ associated
with source bit $\Ssca_i$.  We reveal the check and associated source
bit in a sequential manner for each $i=1, \ldots, \numbit$, and so
define a sequence of random variables $\{U_0, U_1, \ldots, U_\numbit
\}$ via $U_0 \mydefn \Exs[\rvdistcom \; | \; \mycodebarldpc]$, and
\begin{eqnarray}
U_i & \mydefn & \Exs \left[ \rvdistcom \, \mid \; \Ssca_1, \ldots,
\Ssca_i, \, \mycodebarldpc \right], \qquad i=1, \ldots, \numbit.
\end{eqnarray}
By construction, we have $U_\topbit = (\rvdistcom \, \mid \,
\mycodebarldpc)$.  Moreover, this sequence satisfies the following
bounded difference property: adding any source bit $\Ssca_i$ and the
associated check in moving from $U_{i-1}$ to $U_i$ can lead to a
(renormalized) change in the minimum distortion of at most $c_i =
1/\topbit$.  Consequently, by applying Azuma's inequality~\cite{Alon},
we have, for any $\epsilon > 0$,
\begin{eqnarray}
\label{EqnSharpConcentrate}
\mprob\left[ \big| (\rvdistcom \, \mid \, \mycodebarbot) -
\Exs[\rvdistcom \; | \; \mycodebarldpc] \big| \geq \epsilon \right] &
\leq &  \exp \left(-\topbit \epsilon^2 \right).
\end{eqnarray}

Next we observe that our assumption~\eqref{EqnInvPolyDecay} of inverse
polynomial decay implies that, for at least one bottom code
$\mycodebarbot$,
\begin{eqnarray}
\label{EqnAtLeastOne}
\mprob[\rvdistcom \leq \distor \; \mid \; \mycodebarbot] \; = \;
\mprob[\Complexnumc > 0 \; \mid \; \mycodebarbot] & \geq & 1/g(n),
\end{eqnarray}
for some subexponential function $g$.  Otherwise, there would exist
some $\alpha > 0$ such that 
\[
\mprob[\Complexnumc > 0 \, \mid \, \mycodebarbot] \leq \exp(-\topbit
\alpha)
\]
for all choices of bottom code $\mycodebarbot$, and taking averages
would violate our assumption~\eqref{EqnInvPolyDecay}.

Finally, we claim that the concentration
result~\eqref{EqnSharpConcentrate} and inverse polynomial
bound~\eqref{EqnAtLeastOne} yield the result.  Indeed, if for some
$\epsilon > 0$, we had $D < \Exs[\rvdistcom \; | \; \mycodebarldpc]
-\epsilon$, then the concentration bound~\eqref{EqnSharpConcentrate}
would imply that the probability
\begin{eqnarray*}
\mprob[\rvdistcom \leq \distor \; \mid \; \mycodebarbot] & \leq &
\mprob[\rvdistcom \leq \Exs[\rvdistcom \; | \; \mycodebarldpc]
-\epsilon \; \mid \; \mycodebarbot] \\
& \leq & \mprob\left[ \big| (\rvdistcom \, \mid \, \mycodebarbot) -
\Exs[\rvdistcom \; | \; \mycodebarldpc] \big| \geq \epsilon \right]
\end{eqnarray*}
decays exponentially, which would contradict the inverse polynomial
bound~\eqref{EqnAtLeastOne} for sufficiently large $\topbit$.  Thus,
we have shown that assumption~\eqref{EqnInvPolyDecay} implies that for
all $\epsilon > 0$, there exists a sufficiently large $\topbit$ and
fixed bottom code $\mycodebarbot$ such that $\Exs[\rvdistcom \; | \;
\mycodebarldpc] \leq \distor + \epsilon$.  If the average over LDGM
codes $\mycodetop$ satisfies this bound, then at least one choice of
LDGM top code must also satisfy it, whence we have established that
there exists a fixed code $\mycodebar$ such that $\Exs[d_n(\Ssca;
\bar{\mycode})] \leq \distor + \epsilon$, as claimed.
\end{proof}

\subsection{Moment analysis}

In order to analyze the probability $\mprob[\Complexnumc > 0]$, we
make use of the moment bounds given in the following elementary lemma:
\blems[Moment methods]
Given any random variable $N$ taking non-negative integer values,
there holds
\begin{equation}
\label{EqnMomentBounds}
\frac{\left(\Exs[N]\right)^2}{\Exs[N^2]} \; \stackrel{(a)}{\leq} \;
\mprob[N> 0] \; \stackrel{(b)}{\leq} \; \Exs[N].
\end{equation}
\elems
\begin{proof}
The upper bound (b) is an immediate consequence of Markov's
inequality, whereas the lower bound (a) follows by applying the
Cauchy-Schwarz inequality~\cite{Grimmett} as follows
\begin{eqnarray*}
\left(\Exs[N] \right)^2 \; = \; \left(\Exs \big[N \: \Ind[N >
    0] \big]\right)^2 & \leq & \Exs[N^2] \; \Exs
\left[\Ind^2[N > 0] \right] \; = \; \Exs[N^2] \; \mprob[N
  > 0].
\end{eqnarray*}
\end{proof}

The remainder of the proof consists in applying these moment bounds to
the random variable $\Complexnumc$, in order to bound the probability
$\mprob[\Complexnumc > 0]$.  We begin by computing the first moment:
\blems[First moment]
\label{LemFirstMoment}
For any code with rate $R$, the expected number of $\distor$-good
codewords scales exponentially as
\begin{eqnarray}
\frac{1}{\numbit} \log \Exs[\Numc] & = & \left[\rate - (1 -
\binent{D}) \right] \; \pm \; o(1).
\end{eqnarray}
\elems

\begin{proof}
First, by linearity of expectation $\Exs[\Numc] =
\sum_{i=0}^{2^{\numbit \rate}-1} \mprob[\cwind{i}(\distor) = 1] \; =
\; 2^{\numbit \rate} \mprob[\cwind{0}(\distor) = 1]$, where we have
used symmetry of the code construction to assert that
$\mprob[\cwind{i}(\distor) = 1] = \mprob[\cwind{0}(\distor) = 1]$ for
all indices $i$.  Now the event $\{\cwind{0}(\distor) = 1\}$ is
equivalent to an i.i.d Bernoulli($\myhalf)$ sequence of length
$\numbit$ having Hamming weight less than or equal to $\distor
\numbit$.  By standard large deviations theory (either Sanov's
theorem~\cite{Cover}, or direct asymptotics of binomial coefficients),
we have
\begin{eqnarray*}
\frac{1}{\numbit} \log \mprob[\cwind{0}(\distor) = 1] & = & 1 -
\binent{\distor} \; \pm \; o(1),
\end{eqnarray*}
which establishes the claim.

\end{proof}
Unfortunately, however, the first moment $\Exs[\Numc]$ need not be
representative of typical behavior of the random variable $\Numc$, and
hence overall distortion performance of the code.  As a simple
illustration, consider an imaginary code consisting of $2^{\numbit
\rate}$ copies of the all-zeroes codeword.  Even for this ``code'', as
long as \mbox{$\rate > 1-\binent{\distor}$,} the expected number of
distortion-$D$ optimal codewords grows exponentially.  Indeed,
although $\Numc = 0$ for almost all source sequences, for a small
subset of source sequences (of probability mass $\approx 2^{-\numbit
\, \left[1-h(D)\right]}$), the random variable $\Numc$ takes on the
enormous value $2^{nR}$, so that the first moment grows exponentially.
However, the average distortion incurred by using this code will be
$\approx 0.5$ for any rate, so that the first moment is entirely
misleading.  In order to assess the representativeness of the first
moment, one needs to ensure that it is of essentially the same order
as the variance, hence the comparison involved in the second moment
bound~\eqref{EqnMomentBounds}(a).

\subsection{Second moment analysis}

\noindent Our analysis of the second moment begins with the following
alternative representation:
\blems
\label{LemSecondMomentSimple}
\begin{equation}
\label{EqnSecondMomentSimple}
\Exs[\Numc^2(\distor)] \; = \; \Exs[\Numc(\distor)] \, \Biggr( 1 +
\Big \{ \sum_{j \neq 0} \mprob[ \cwind{j}(\distor) =1 \, \mid \,
\cwind{0}(\distor) =1] \Big\} \Biggr).
\end{equation}
\elems
\noindent Based on this lemma, proved in
Appendix~\ref{AppSecondMomentSimple}, we see that the key quantity to
control is the conditional probability $\mprob[ \cwind{j}(\distor) =1
\, \mid \, \cwind{0}(\distor) =1]$.  It is this \emph{overlap
probability} that differentiates the low-density codes of interest
here from the unstructured codebooks used in classical random coding
arguments.\footnote{In the latter case, codewords are chosen
independently from some ensemble, so that the overlap probability is
simply equal to $\mprob[ \cwind{j}(\distor) =1]$.  Thus, for the
simple case of unstructured random coding, the second moment bound
actually provides the converse to Shannon's rate-distortion theorem
for the symmetric Bernoulli source.}  For a low-density graphical
code, the dependence between the events $\{\cwind{j}(\distor) =1 \}$
and $\{\cwind{0}(\distor) =1 \}$ requires some analysis.

Before proceeding with this analysis, we require some definitions.
Recall our earlier definition~\eqref{EqnDefnAvWtEnum} of the average
weight enumerator associated with an $(\vdeg, \lowcdeg)$ LDPC code,
denoted by $\AvWtEnum{\numinfobit}(\mywei)$.  Moreover, let us define
for each $\mywei \in [0,1]$ the probability
\begin{eqnarray}
\Qprob(\mywei; \distor) & \mydefn & \mprob \left[ \|X(\mywei) \oplus
\Ssca\|_1 \leq \distor \numbit \; \mid \; \|\Ssca \|_1 \leq \distor
\numbit \right],
\end{eqnarray}
where the quantity $\Codebit(\mywei) \in \{0,1\}^\numcodebit$ denotes
a randomly chosen codeword, conditioned on its underlying
length-$\numinfobit$ information sequence having Hamming weight
$\lceil \mywei \numinfobit \rceil$.  As shown in
Lemma~\ref{LemUpperBer} (see Appendix~\ref{AppUpperBer}), the random
codeword $\Codebit(\mywei)$ has i.i.d. Bernoulli elements with
parameter
\begin{eqnarray}
\label{EqnDefnDelfun}
\delfun{\mywei} & = & \onehalf \, \biggr[ 1 - (1 - 2 \,
\mywei)^{\topdeg} \biggr].
\end{eqnarray}

With these definitions, we now break the sum on the RHS of
equation~\eqref{EqnSecondMomentSimple} into $\numinfobit$ terms,
indexed by $t = 1,2, \ldots, \numinfobit$, where term $t$ represents
the contribution of a given non-zero information sequence $\infobit
\in \{0,1\}^\numinfobit$ with (Hamming) weight $t$.  Doing so yields
\begin{eqnarray*}
\sum_{j \neq 0} \mprob[ \cwind{j}(\distor) =1 \, \mid \,
\cwind{0}(\distor) =1] & = & \sum_{\mytvar=1}^\numinfobit
\AvWtEnum{\numinfobit}(\mytvar/\numinfobit) \,
\Qprob(\mytvar/\numinfobit; \distor) \\
& \leq & \numinfobit \max_{1 \leq \mytvar \leq \numinfobit} \left \{
  \AvWtEnum{\numinfobit}(\mytvar/\numinfobit) \;
  \Qprob(\mytvar/\numinfobit; \distor) \right \}\\
& \leq & \numinfobit \max_{\mywei \in [0,1]} \left \{
\AvWtEnum{\numinfobit}(\mywei) \; \Qprob(\mywei; \distor) \right \}.
\end{eqnarray*}
Consequently, we need to control both the LDPC weight enumerator
$\AvWtEnum{\numinfobit}(\mywei)$ and the probability $\Qprob(\mywei;
\distor)$ over the range of possible fractional weights $\mywei \in
[0,1]$.

\subsection{Bounding the overlap probability}

The following lemma, proved in Appendix~\ref{AppQprob}, provides a
large deviations bound on the probability $\Qprob(\mywei; \distor)$.
\blems
\label{LemQprob}
For each $\mywei \in [0,1]$, we have 
\begin{eqnarray}
\label{EqnKeyFuncUpper}
\frac{1}{\topbit} \log \Qprob(\mywei; \distor) & \leq &
\KeyFunc(\IndBer{\mywei}; \distor) + o(1), 
\end{eqnarray}
where for each $\tmpvar \in (0,\myhalf]$ and $\distor \in (0,
\myhalf]$, the error exponent is given by
\begin{eqnarray}
\label{EqnDefnKeyFunc}
\KeyFunc(\tmpvar; \distor) & \mydefn & \distor \log \left[(1-\tmpvar)
e^{\lamstar} + \tmpvar\right] + (1-\distor) \log \left[ (1-\tmpvar) +
\tmpvar e^{\lamstar} \right] - \lamstar\distor. \qquad
\end{eqnarray}
Here $\lamstar \mydefn \log \biggr [ \frac{-b + \sqrt{b^2 - 4 a c} }{2
a} \biggr]$, where $a \mydefn \tmpvar \, (1-\tmpvar) \, (1-\distor)$,
$b \mydefn (1-2 \distor) \tmpvar^2$, and $c \mydefn -\tmpvar \,
(1-\tmpvar) \, \distor$.  \elems

\begin{figure}[h]
\begin{center}
\begin{tabular}{cc}
\widgraph{0.5\textwidth}{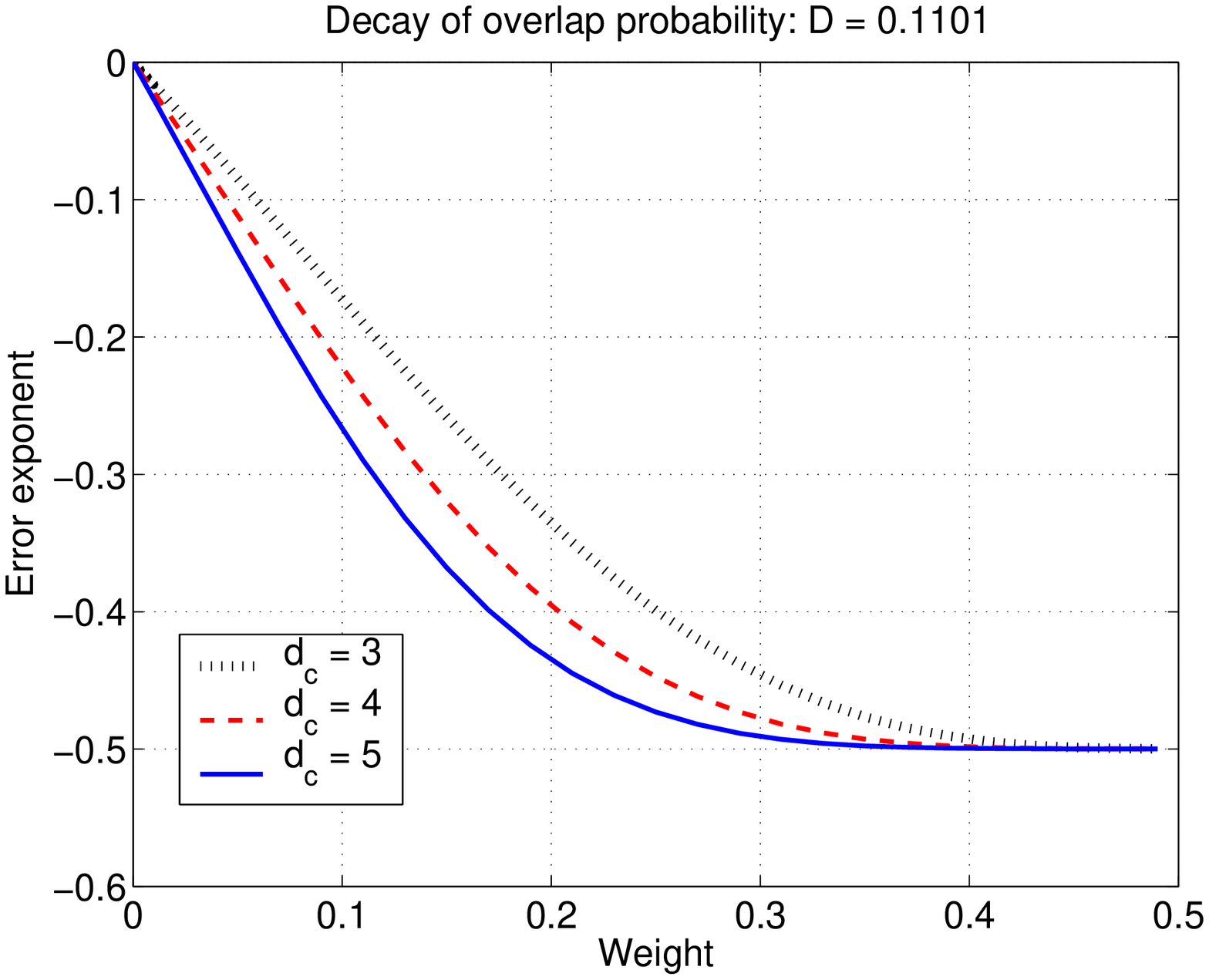} &
\widgraph{0.5\textwidth}{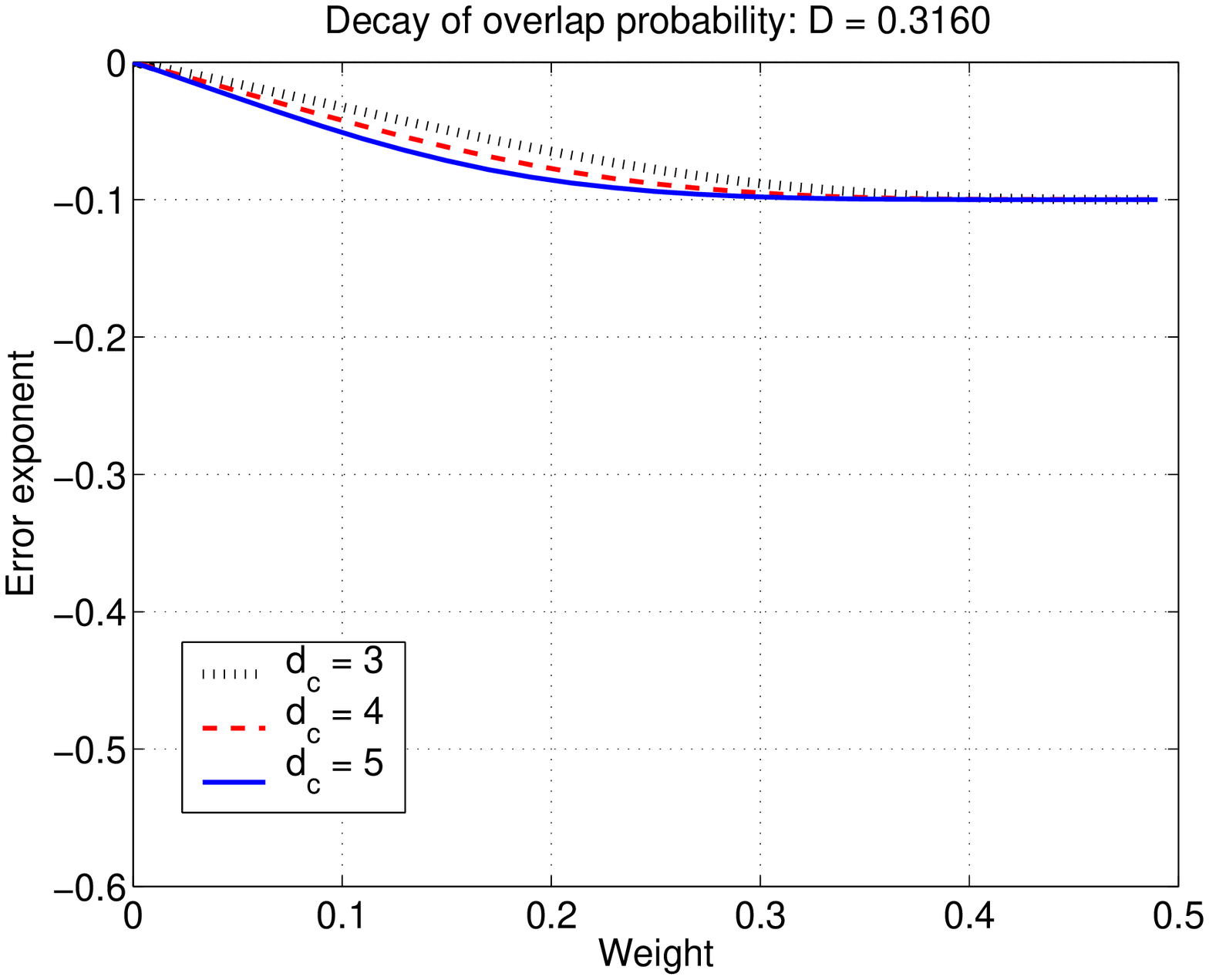} \\
(a) & (b)
\end{tabular}
\caption{Plot of the upper bound~\eqref{EqnKeyFuncUpper} on the
overlap probability $\frac{1}{\topbit} \log \Qprob(\mywei; \distor)$
for different choices of the degree $\topdeg$, and distortion
probabilities. (a) Distortion $\distor = 0.1100$.  (b) Distortion
$\distor = 0.3160$.}
\label{FigQprob}
\end{center}
\end{figure}

In general, for any $\distor \in (0, \myhalf]$, the function
$\KeyFunc(\, \cdot \,; \distor)$ has the following properties.  At
\mbox{$\tmpvar = 0$,} it achieves its maximum $\KeyFunc(0 \, ;
\distor) = 0$, and then is strictly decreasing on the interval $(0,
\myhalf]$, approaching its minimum value $-\left[1-\binent{\distor}
\right]$ as $\tmpvar \rightarrow \myhalf$.  Figure~\ref{FigQprob}
illustrates the form of the function $\KeyFunc(\delfun{\omega};
\distor)$ for two
different values of distortion $\distor$, and for degrees $\topdeg \in
\{3,4,5\}$.  Note that increasing $\topdeg$ causes
$\KeyFunc(\delfun{\omega};\distor)$ to approach its minimum
$-[1-\binent{\distor}]$ more rapidly.

\begin{figure}[h]
\begin{center}
\begin{tabular}{cc}
\widgraph{0.5\textwidth}{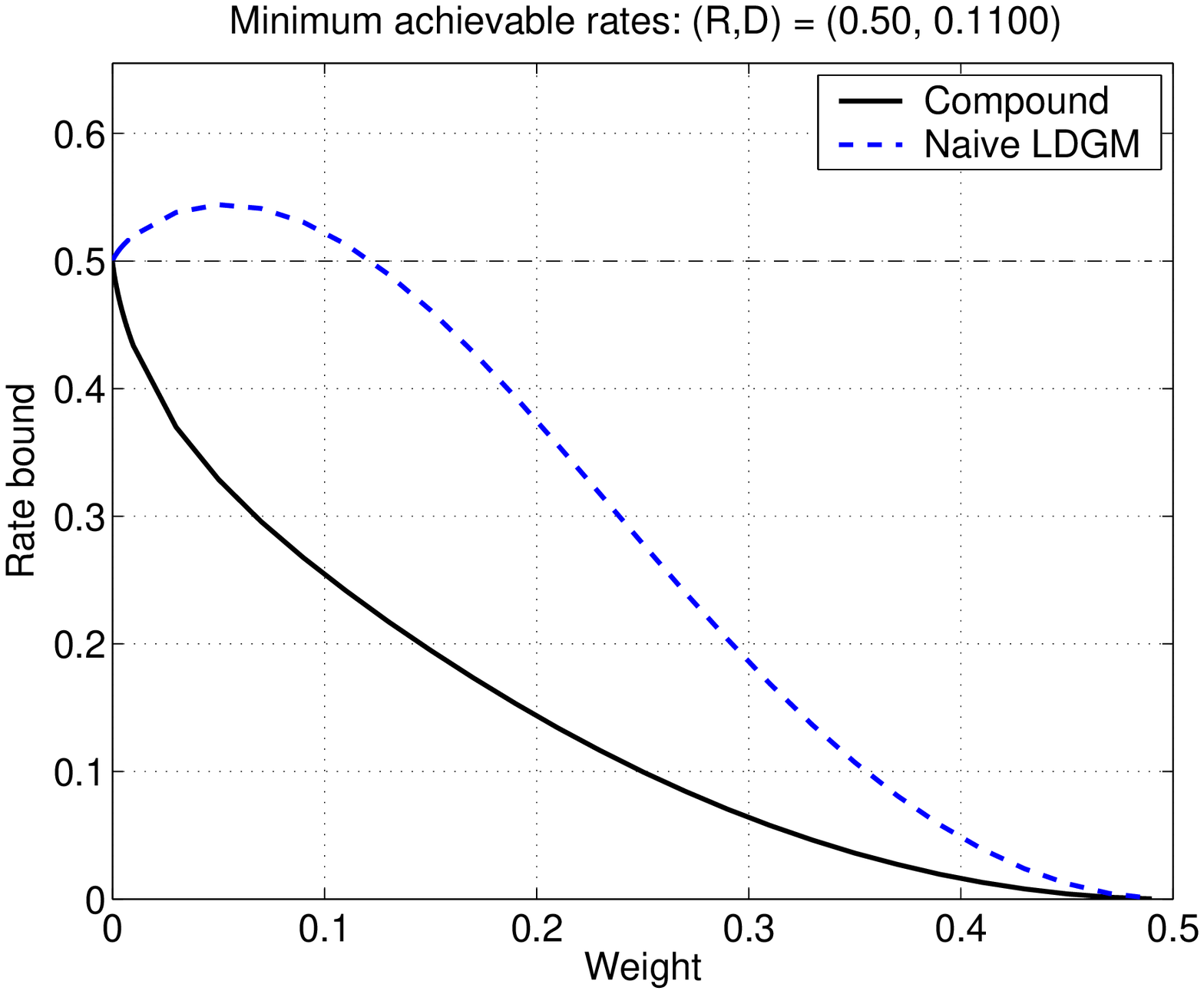} &
\widgraph{0.515\textwidth}{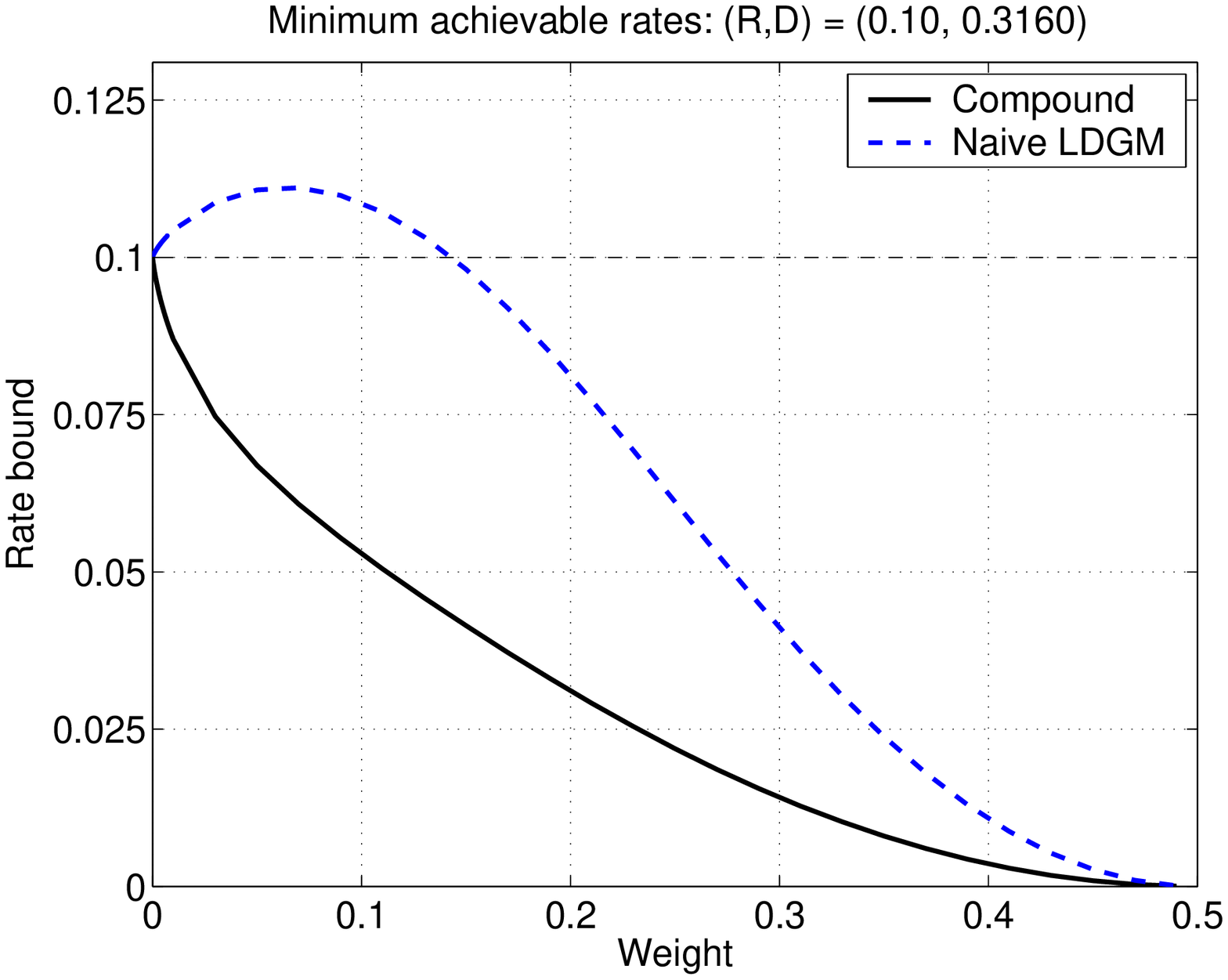} \\
(a) & (b)
\end{tabular}
\caption{Plot of the function defining the lower
bound~\eqref{EqnTheoremBound} on the minimum achievable rate for a
specified distortion.  Shown are curves with LDGM top degree $\topdeg
= 4$, comparing the uncoded case (no bottom code, dotted curve) to a
bottom $(4,6)$ LDPC code (solid line).  (a) Distortion $\distor =
0.1100$.  (b) Distortion $\distor = 0.3160$.}
\label{FigThmBound}
\end{center}
\end{figure}

We are now equipped to establish the form of the effective
rate-distortion function for any compound LDGM/LDPC
ensemble. Substituting the alternative form of $\Exs[\Numc^2]$ from
equation~\eqref{EqnSecondMomentSimple} into the second moment lower
bound~\eqref{EqnMomentBounds} yields
\begin{eqnarray}
\frac{1}{\numbit} \log \Prob[\Numc(\distor) > 0] & \geq &
\frac{1}{\numbit} \left[ \log \Exs[\Numc(\distor)] - \log \left \{1
+\sum_{j \neq 0} \Prob[\cwind{j}(\distor) =1 \, \mid \;
\cwind{0}(\distor) = 1] \right \} \right] \nonumber \\
& \geq & \ratetot - \left(1- \binent{\distor} \right) - \max_{\mywei
\in [0,1]} \left \{ \frac{1}{\numbit} \log
\AvWtEnum{\numinfobit}(\mywei) + \frac{1}{\numbit} \log \Qprob(\mywei;
\distor) \right \} - o(1) \nonumber \\
\label{EqnLastStep}
& \geq & \ratetot - \left(1- \binent{\distor} \right) - \max_{\mywei
\in [0,1]} \left\{ \ratetot \; \frac{1}{\rateldpc} \frac{\log
\AvWtEnum{\numinfobit}(\mywei)}{\numinfobit} + \KeyFunc(\
\IndBer{\mywei}, \distor) \right \} - o(1), \qquad
\end{eqnarray}
where the last step follows by applying the upper bound on $\Qprob$
from Lemma~\ref{LemQprob}, and the relation $\numinfobit = \rateldgm
\numbit = \frac{\ratetot}{\rateldpc} \numbit$. Now letting
$\BouWtEnum(\mywei; \vdeg, \lowcdeg)$ be any upper bound on the log
of average weight enumerator $\frac{\log
\AvWtEnum{\numinfobit}(\mywei)}{\numinfobit}$, we can then conclude
that \mbox{$\frac{1}{\numbit} \log \Prob[\Numc(\distor) > 0]$} is
asymptotically non-negative for all rate-distortion pairs $(\myrate,
\distor)$ satisfying
\begin{eqnarray}
\label{EqnTheoremBound}
\ratetot & \geq & \max_{\mywei \in [0,1]} \left[\frac{1 -
\binent{\distor} + \KeyFunc(\IndBer{\mywei},
\distor)}{1-\frac{\BouWtEnum(\mywei; \vdeg, \lowcdeg)}{\rateldpc}}
\right].
\end{eqnarray}
\noindent Figure~\ref{FigThmBound} illustrates the behavior of the RHS
of equation~\eqref{EqnTheoremBound}, whose maximum defines the
effective rate-distortion function, for the case of LDGM top degree
$\topdeg = 4$.  Panels (a) and (b) show the cases of distortion
$\distor = 0.1100$ and $\distor = 0.3160$ respectively, for which the
respective Shannon rates are $R = 0.50$ and $R = 0.10$.  Each panel
shows two plots, one corresponding the case of uncoded information
bits (a naive LDGM code), and the other to using a rate $\rateldpc =
2/3$ LDPC code with degrees $(\vdeg, \cdeg) = (4,6)$.  In all cases,
the minimum achievable rate for the given distortion is obtained by
taking the maximum for $\mywei \in [0,0.5]$ of the plotted function.
For any choices of $\distor$, the plotted curve is equal to the
Shannon bound $\ratesha = 1-\binent{\distor}$ at $\mywei = 0$, and
decreases to $0$ for $\mywei = \myhalf$.

Note the dramatic difference between the uncoded and compound
constructions (LDPC-coded). In particular, for both settings of the
distortion ($\distor =0.1100$ and $\distor = 0.3160$), the uncoded
curves rise from their initial values to maxima \emph{above} the
Shannon limit (dotted horizontal line).  Consequently, the minimum
required rate using these constructions lies strictly above the
Shannon optimum.  The compound construction curves, in contrast,
decrease monotonically from their maximum value, achieved at $\mywei =
0$ and corresponding to the Shannon optimum.  In the following
section, we provide an analytical proof of the fact that for any
distortion $\distor \in [0, \myhalf)$, it is always possible to choose
finite degrees such that the compound construction achieves the
Shannon optimum.

\subsection{Finite degrees are sufficient}
\label{SecFiniteDegrees}
In order to complete the proof of Theorem~\ref{ThmSource}, we need to
show that for all rate-distortion pairs $(\myrate, \distor)$
satisfying the Shannon bound, there exist LDPC codes with finite
degrees $(\vdeg, \lowcdeg)$ and a suitably large but finite top degree
$\topdeg$ such that the compound LDGM/LDPC construction achieves the
specified $(\myrate, \distor)$.

Our proof proceeds as follows.  Recall that in moving from
equation~\eqref{EqnLastStep} to equation~\eqref{EqnTheoremBound}, we
assumed a bound on the average weight enumerator
$\AvWtEnum{\numinfobit}$ of the form
\begin{eqnarray}
\label{EqnLDPCAss}
\frac{1}{\numinfobit} \log \AvWtEnum{\numinfobit}(\mywei) & \leq &
\AsympWtEnum(\mywei; \vdeg, \lowcdeg) + o(1).
\end{eqnarray}
For compactness in notation, we frequently write
$\AsympWtEnum(\mywei)$, where the dependence on the degree pair
$(\vdeg, \lowcdeg)$ is understood implicitly.  In the following
paragraph, we specify a set of conditions on this bounding function
$\AsympWtEnum$, and we then show that under these conditions, there
exists a finite degree $\topdeg$ such that the compound construction
achieves specified rate-distortion point.  In
Appendix~\ref{AppLDPCSuffice}, we then prove that the weight
enumerator of standard regular LDPC codes satisfies the assumptions
required by our analysis.

\paragraph{Assumptions on weight enumerator bound}
We require that our bound $\AsympWtEnum$ on the weight enumerator
satisfy the following conditions:
\begin{enumerate}
\item[{\bf{A1:}}] the function $\AsympWtEnum$ is symmetric around
$\myhalf$, meaning that $\AsympWtEnum(\mywei) =
\AsympWtEnum(1-\mywei)$ for all $\mywei \in [0,1]$.
\item[{\bf{A2:}}] the function $\AsympWtEnum$ is twice differentiable on
$(0,1)$ with $\AsympWtEnum'(\myhalf) = 0$ and $\AsympWtEnum''(\myhalf)
< 0$.
\item[{\bf{A3:}}] the function $\AsympWtEnum$ achieves its unique
optimum at $\mywei = \myhalf$, where $\AsympWtEnum(\myhalf) =
\rateldpc$.
\item[{\bf{A4:}}] there exists some $\epslow > 0$ such that
$\AsympWtEnum(\mywei) < 0$ for all $\mywei \in (0,\epslow)$, meaning
that the ensemble has linear minimum distance.
\end{enumerate}

In order to establish our claim, it suffices to show that for all
$(\myrate, \distor)$ such that $\myrate > 1 - \binent{\distor}$, there
exists a finite choice of $\topdeg$ such that
\begin{eqnarray}
\label{EqnToBeProven}
\max_{\mywei \in [0,1]} \left\{ \underbrace{\ratetot \;
 \frac{\AsympWtEnum(\mywei)}{\rateldpc} + \KeyFunc(\IndBer{\mywei},
 \distor)} \right \} & \leq & \myrate - \left[1 - \binent{\distor}
 \right] \; \mydefn \; \Delta \\
\ProofFun(\mywei; \topdeg) \qquad \qquad \qquad & & \nonumber
\end{eqnarray}
Restricting to even $\topdeg$ ensures that the function $\KeyFunc$ is
symmetric about $\mywei = \myhalf$; combined with assumption A2, this
ensures that $\ProofFun$ is symmetric around $\myhalf$, so that we may
restrict the maximization to $[0, \myhalf]$ without loss of
generality.  Our proof consists of the following steps:
\begin{enumerate}
\item[(a)] We first prove that there exists an $\epslow > 0$,
independent of the choice of $\topdeg$, such that
\mbox{$\ProofFun(\mywei; \topdeg) \leq \Delta$} for all $\mywei \in [0,
\epslow]$.
\item[(b)] We then prove that there exists $\epsup > 0$, again
independent of the choice of $\topdeg$, such that $\ProofFun(\mywei;
\topdeg) \leq \Delta$ for all $\mywei \in [\myhalf - \epsup,
\myhalf]$.
\item[(c)] Finally, we specify a sufficiently large but finite degree
$\topdeg^*$ that ensures the condition \mbox{$\ProofFun(\mywei;
\topdeg^*) \leq \Delta$} for all $\mywei \in [\epslow, \epsup]$.
\end{enumerate}

\subsubsection{Step A}
By assumption A4 (linear minimum distance), there exists some $\epslow
> 0$ such that $\AsympWtEnum(\mywei) \leq 0$ for all $\mywei \in [0,
  \epslow]$.  Since $\KeyFunc(\IndBer{\mywei}; \distor) \leq 0$ for
all $\mywei$, we have $\ProofFun(\mywei; \topdeg) \leq 0 < \Delta$ in
this region. Note that $\epslow$ is independent of $\topdeg$, since it
specified entirely by the properties of the bottom code.

\subsubsection{Step B}

For this step of the proof, we require the following lemma on the
properties of the function $\KeyFunc$:
\blems
\label{LemFbound}
For all choices of even degrees $\topdeg \geq 4$, the function
$G(\mywei; \topdeg) = \KeyFunc(\IndBer{\mywei}, \distor)$ is
differentiable in a neighborhood of $\mywei = \myhalf$, with
\begin{equation}
G(\myhalf; \topdeg)  =  - \left[ 1 - \binent{\distor} \right], \qquad
G'(\myhalf; \topdeg)  =  0, \qquad \mbox{and}  \qquad
G''(\myhalf; \topdeg) =  0.
\end{equation}
\elems
See Appendix~\ref{AppFbound} for a proof of this claim.  Next observe
that we have the uniform bound $G(\mywei; \topdeg) \leq G(\mywei; 4)$
for all $\topdeg \geq 4$ and $\mywei \in [0, \myhalf]$.  This follows
from the fact that $\KeyFunc(u; \distor)$ is decreasing in $u$, and
that $\delta^*(\mywei; 4) \leq \delta^*(\mywei; \topdeg)$ for all
$\topdeg \geq 4$ and $\mywei \in [0, \myhalf]$.  Since $\AsympWtEnum$
is independent of $\topdeg$, this implies that $\ProofFun(\mywei;
\topdeg) \leq \ProofFun(\mywei; 4)$ for all $\mywei \in [0,\myhalf]$.
Hence it suffices to set $\topdeg = 4$, and show that
$\ProofFun(\mywei; 4) \leq \Delta$ for all $\mywei \in [\myhalf -
\epsup, \myhalf]$.  Using Lemma~\ref{LemFbound}, Assumption A2
concerning the derivatives of $\AsympWtEnum$, and Assumption A4 (that
$\AsympWtEnum(\myhalf) = \rateldpc$), we have
\begin{eqnarray*}
\ProofFun(\myhalf; 4) & = & \ratetot - \left[1-\binent{\distor}
\right] \; = \; \Delta, \\
\ProofFun'(\myhalf; 4) & = & \frac{\ratetot \;
\AsympWtEnum'(\myhalf)}{\rateldpc} + G'(\myhalf; 4) \; = \; 0, \qquad
\mbox{and} \\
\ProofFun''(\myhalf; 4) & = & \frac{\ratetot \;
\AsympWtEnum''(\myhalf)}{\rateldpc} + G''(\myhalf; 4) \; = \;
\frac{\ratetot \; \AsympWtEnum''(\myhalf)}{\rateldpc} < 0.
\end{eqnarray*}
By the continuity of $\ProofFun''$, the second derivative remains
negative in a region around $\myhalf$, say for all $\mywei \in
[\myhalf - \epsup, \myhalf]$ for some $\epsup > 0$.  Then, for all
$\mywei \in [\myhalf - \epsup, \myhalf]$, we have for some $\myweibar
\in [\mywei, \myhalf]$ the second order expansion
\begin{eqnarray*}
\ProofFun(\mywei; 4) & = & \ProofFun(\myhalf; 4) + \ProofFun'(\myhalf;
4) (\mywei - \myhalf) + \frac{1}{2} \ProofFun(\myweibar; 4) \,
\left(\mywei - \myhalf \right)^2 \\
& = & \Delta + \frac{1}{2} \ProofFun(\myweibar; 4) \, \left(\mywei - \myhalf
\right)^2 \; \leq \;  \Delta.
\end{eqnarray*}
Thus, we have established that there exists an $\epsup > 0$,
independent of the choice of $\topdeg$, such that for all even
$\topdeg \geq 4$, we have
\begin{equation}
\ProofFun(\mywei; \topdeg) \; \leq \; \ProofFun(\mywei, 4) \; \leq \;
\Delta \qquad \mbox{for all $\mywei \in [\myhalf - \epsup, \myhalf]$.}
\end{equation}

\subsubsection{Step C}

Finally, we need to show that $\ProofFun(\mywei; \topdeg) \leq \Delta$
for all $\mywei \in [\epslow, \epsup]$.  From assumption A3 and the
continuity of $\AsympWtEnum$, there exists some $\rho(\epsup) > 0$
such that
\begin{eqnarray}
\label{EqnBouBou}
\BouWtEnum(\mywei) & \leq & \rateldpc \; \left[1 - \rho(\epsup)\right]
\qquad \mbox{for all $\mywei \leq \myhalf - \epsup$.}
\end{eqnarray}
From Lemma~\ref{LemFbound}, $\lim_{u \rightarrow \myhalf} \KeyFunc(u;
\distor) = \KeyFunc(\myhalf;\distor) \; = \; -
\left[1-\binent{\distor} \right]$.  Moreover, as $\topdeg \rightarrow
+\infty$, we have $\IndBer{\epslow} \rightarrow \myhalf$.  Therefore,
for any $\epsilon_3 > 0$, there exists a finite degree $\topdeg^*$
such that
\begin{eqnarray*}
\KeyFunc(\delta^*(\epslow; \topdeg^*);\distor) & \leq &
-\left[1-\binent{\distor}\right] + \epsilon_3.
\end{eqnarray*}
Since $\KeyFunc$ is non-increasing in $\mywei$, we have
$\KeyFunc(\delta^*(\mywei; \topdeg^*); \distor) \leq
-\left[1-\binent{\distor}\right] + \epsilon_3$ for all $\mywei \in
[\epslow, \epsup]$.  Putting together this bound with the earlier
bound~\eqref{EqnBouBou} yields that for all $\mywei \in [\epslow,
\epsup]$:
\begin{eqnarray*}
\ProofFun(\mywei; \topdeg) & = & \ratetot \;
 \frac{\AsympWtEnum(\mywei)}{\rateldpc} + \KeyFunc(\delta^*(\mywei;
 \topdeg^*), \distor) \\
& \leq & \rate \left[1 - \rho(\epsup) \right]
-\left[1-\binent{\distor}\right] + \epsilon_3 \\
& = & \left \{ \rate -\left[1-\binent{\distor}\right] \right \} +
\left(\epsilon_3 - \rate \rho(\epsup) \right) \\
& = & \Delta + \left(\epsilon_3 - \rate \rho(\epsup) \right)
\end{eqnarray*}
Since we are free to choose $\epsilon_3 > 0$, we may set $\epsilon_3 =
\frac{\rate \rho(\epsup)}{2}$ to yield the claim.


\section{Proof of channel coding optimality}
\label{SecChannelOptimal}

In this section, we turn to the proof of the previously stated
Theorem~\ref{ThmChannel}, concerning the channel coding optimality of
the compound construction.

  If the codeword $\codebit \in \{0,1\}^\numbit$ is transmitted, then
the receiver observes $\Recbit = \codebit \oplus \Wnoise$, where
$\Wnoise$ is a $\myber(\channoise)$ random vector.  Our goal is to
bound the probability that maximum likelihood (ML) decoding fails
where the probability is taken over the randomness in both the channel
noise and the code construction.  To simplify the analysis, we focus
on the following sub-optimal (non-ML) decoding procedure.  Let
$\myeps$ be any non-negative sequence such that $\myeps/\topbit
\rightarrow 0$ but $\myeps^2/\topbit \rightarrow +\infty$---say for
instance, $\myeps = \topbit^{2/3}$.
\bdes[Decoding Rule:]
\label{DefnSubopt}
With the threshold $d(\topbit) \mydefn p \topbit + \myeps$, decode to
codeword $\codebit_i$ $\iff$ \mbox{$\|\codebit_i \oplus \Recbit \|_1
\leq d(\topbit)$,} and no other codeword is within $d(\topbit)$ of
$\Recbit$.  \edes
\noindent The extra term $\myeps$ in the threshold $d(\topbit)$ is
chosen for theoretical convenience.
Using the following two lemmas, we establish that this procedure has
arbitrarily small probability of error, whence ML decoding (which is
at least as good) also has arbitrarily small error probability.
\begin{lems}
Using the suboptimal procedure specified in the
definition~\eqref{DefnSubopt}, the probability of decoding error
vanishes asymptotically provided that
\begin{equation}
\label{EqnErrExponent}
\rateldgm \; \AsympWtEnum(\mywei) - \rent{\channoise}{\delfun{\mywei}
\ast \channoise} \; < \; 0 \qquad \mbox{ for all $\mywei \in
(0,\myhalf]$,}
\end{equation}
where $\AsympWtEnum$ is any function bounding the average weight
enumerator as in equation~\eqref{EqnLDPCAss}.
\end{lems}
\begin{proof}
Let $N = 2^{\topbit \ratetot} = 2^{\midbit \rateldpc}$ denote the
total number of codewords in the joint LDGM/LDPC code.  Due to the
linearity of the code construction and symmetry of the decoding
procedure, we may assume without loss of generality that the all zeros
codeword $\zeroes$ was transmitted (\ie, $\codebit = \zeroes$).  In
this case, the channel output is simply $\Recbit = \Wnoise$ and so our
decoding procedure will fail if and only if one the following two
conditions holds:
\begin{enumerate}
\item[(i)] either $\|\Wnoise \|_1 > d(\topbit)$, or
\item[(ii)] there exists a sequence of information bits $\infobit \in
\{0,1\}^\midbit$ satisfying the parity check equation $\Parmat
\infobit = 0$ such that the codeword $\Genmat \infobit$ satisfies
$\|\Genmat \infobit \oplus \Wnoise\|_1 \leq d(\topbit)$.
\end{enumerate}
Consequently, using the union bound, we can upper bound the
probability of error as follows:
\begin{equation}
\label{EqnErr}
p_{err} \leq \Prob [\|\Wnoise\|_1 > d(\topbit) ] + \sum_{i=2}^N \Prob
\big[\|\Genmat \infobit^i \oplus \Wnoise\|_1 \leq d(\topbit)
\big]. \quad
\end{equation}
Since $\Exs[\|\Wnoise\|_1] = \channoise \numbit$, we may apply
Hoeffdings's inequality~\cite{Devroye} to conclude that
\begin{eqnarray}
\Prob [\|\Wnoise\|_1 > d(\topbit) ] & \leq & 2 \exp\left(-2
\frac{\myeps^2}{\topbit} \right)  \; \rightarrow \; 0
\end{eqnarray}
by our choice of $\myeps$.  Now focusing on the second term, let us
rewrite it as a sum over the possible Hamming weights $\ell = 1, 2,
\ldots, \midbit$ of information sequences (i.e., $\|\infobit\|_1 =
\ell$) as follows:
\begin{eqnarray*}
\sum_{i=2}^N \Prob \big[\|\Genmat \infobit^i \oplus \Wnoise\|_1 \leq
d(\topbit) \big] & = & \sum_{\ell=1}^\midbit
\AvWtEnum{\numinfobit}(\frac{\ell}{\midbit}) \; \Prob\big[\|\Genmat
\infobit \oplus \Wnoise\|_1 \geq d(\topbit) \; \big | \;
\|\infobit\|_1 = \ell \big],
\end{eqnarray*}
where we have used the fact that the (average) number of information
sequences with fractional weight $\ell/\midbit$ is given by the LDPC
weight enumerator $\AvWtEnum{\numinfobit}(\frac{\ell}{\midbit})$.
Analyzing the probability terms in this sum, we note
Lemma~\ref{LemUpperBer} (see Appendix~\ref{AppUpperBer}) guarantees
that $\Genmat \infobit$ has
i.i.d. $\myber(\delfun{\frac{\ell}{\midbit}})$ elements, where
$\delfun{\, \cdot \,}$ was defined in equation~\eqref{EqnDefnDelfun}.
Consequently, the vector $\Genmat \infobit \oplus \Wnoise$ has
i.i.d. $\myber(\inducedWeight{\frac{\ell}{\midbit}} \ast \channoise)$
elements.  Applying Sanov's theorem~\cite{Cover} for the special case
of binomial variables yields that for any information bit sequence
$\infobit$ with $\ell$ ones, we have
\begin{eqnarray}
\label{EqnChanProbUpper}
\Prob\big[\|\Genmat \infobit \oplus \Wnoise\|_1 \geq d(\topbit) \;
\big | \; \|\infobit\|_1 = \ell \big] & \leq & f(\topbit) 2^{-\topbit
\rent{\channoise}{\inducedWeight{\frac{\ell}{\midbit}} \ast
\channoise}},
\end{eqnarray}
for some polynomial term $f(n)$.  We can then upper bound the second
term in the error bound~\eqref{EqnErr} as
\begin{eqnarray*}
\sum_{i=2}^N \Prob \big[\|\Genmat \infobit^i \oplus \Wnoise\|_1 \leq
d(\topbit) \big] & \leq & f(\midbit) \; \exp \Biggr \{ \max_{1 \leq
\ell \leq \midbit} \Big[ \midbit \AsympWtEnum(\frac{\ell}{\midbit}) +
o(\midbit) -\topbit
\rent{\channoise}{\inducedWeight{\frac{\ell}{\midbit}} \ast
\channoise} \Big ] \Biggr \},
\end{eqnarray*}
where we have used equation~\eqref{EqnChanProbUpper}, as well as the
assumed upper bound~\eqref{EqnLDPCAss} on $\AvWtEnum{\numinfobit}$ in
terms of $\AsympWtEnum$.  Simplifying further, we take logarithms and
rescale by $\midbit$ to assess the exponential rate of decay, thereby
obtaining
\begin{eqnarray*}
\frac{1}{\midbit} \log \sum_{i=2}^N \Prob \big[\|\Genmat \infobit^i
\oplus \Wnoise\|_1 \leq d(\topbit) \big] & \leq & \max_{1 \leq \ell
\leq \midbit} \Big[ \AsympWtEnum(\frac{\ell}{\midbit}) -
\frac{1}{\rateldgm}
\rent{\channoise}{\inducedWeight{\frac{\ell}{\midbit}} \ast
\channoise} \Big ] + o(1) \\
& \leq & \max_{\mywei \in [0,1]} \Big[ \AsympWtEnum(\mywei) -
\frac{1}{\rateldgm} \rent{\channoise}{\inducedWeight{\mywei} \ast
\channoise} \Big ] + o(1),
\end{eqnarray*}
and establishing the claim.
\end{proof}

\begin{lems} 
\label{lem:easy_lemma}
For any $\channoise \in (0,1)$ and total rate $\ratetot \mydefn
\rateldgm \, \rateldpc \; < \; 1 -\binent{\channoise}$, there exist
finite choices of the degree triplet ($\topdeg, \vdeg, \lowcdeg)$ such
that \eqref{EqnErrExponent} is satisfied.
\end{lems}
\begin{proof}
For notational convenience, we define
\begin{eqnarray}
\label{EqnDefnErrFun}
\ErrFun(\mywei) & \mydefn & \rateldgm 
\AsympWtEnum(\mywei) -
\rent{\channoise}{\delfun{\mywei} \ast \channoise}.
\end{eqnarray}
First of all, it is known~\cite{Gallager63} that a regular LDPC code
with rate $\rateldpc = \frac{\vdeg}{\lowcdeg} < 1$ and $\vdeg \geq 3$
has linear minimum distance.  More specifically, there exists a
threshold $\ldpcthresh = \ldpcthresh(\vdeg, \cdeg)$ such that
$\AsympWtEnum(\mywei) \leq 0$ for all $\mywei \in [0, \ldpcthresh]$.
Hence, since $\AsympWtEnum(\mywei) - \rent{\channoise}{\delfun{\mywei}
\ast \channoise} \geq 0$ for all $\mywei \in (0,1)$, for $\mywei \in
(0, \ldpcthresh]$, we have $\ErrFun(\mywei) < 0$.

Turning now to the interval $[\ldpcthresh, \myhalf]$, consider the
function
\begin{eqnarray}
\label{EqnDefnErrFunTil}
\ErrFunTil(\mywei) & \mydefn & \ratecom \binent{\mywei} -
\rent{\channoise}{\delfun{\mywei} \ast \channoise}.
\end{eqnarray}
Since $\AsympWtEnum(\mywei) \leq \rateldpc \binent{\mywei}$, we have
$\ErrFun(\mywei) \leq \ErrFunTil(\mywei)$, so that it suffices to
upper bound $\ErrFunTil$.  Observe that $\ErrFunTil(\myhalf) =
\ratecom - (1 - \binent{\channoise}) < 0$ by assumption. Therefore, it
suffices to show that, by appropriate choice of $\topdeg$, we can
ensure that $\ErrFunTil(\mywei) \leq \ErrFunTil(\myhalf)$.  Noting
that $\ErrFunTil$ is infinitely differentiable, calculating
derivatives yields $\ErrFunTil'(\myhalf) = 0$ and
$\ErrFunTil''(\myhalf) < 0$.  (See Appendix~\ref{AppErrFunTilDeriv}
for details of these derivative calculations.)  Hence, by  second
order Taylor series expansion around $\mywei = \myhalf$, we obtain
\begin{eqnarray*}
\ErrFunTil(\mywei) & = & \ErrFunTil(\myhalf) + \frac{1}{2}
\ErrFunTil''(\bar{\mywei}) (\mywei - \myhalf)^2,
\end{eqnarray*}
where $\bar{\mywei} \in [\mywei, \myhalf]$.  By continuity of
$\ErrFunTil''$, we have $\ErrFunTil''(\mywei) < 0$ for all $\mywei$ in
some neighborhood of $\myhalf$, so that the Taylor series expansion
implies that $\ErrFunTil(\mywei) \leq \ErrFunTil(\myhalf)$ for
all $\mywei$ in some neighborhood, say $(\lowtay, \myhalf]$.

It remains to bound $\ErrFunTil$ on the interval $[\ldpcthresh,
\lowtay]$.  On this interval, we have $\ErrFunTil(\mywei) \leq
\ratecom \binent{\lowtay} - \rent{\channoise}{\delfun{\ldpcthresh}
\ast \channoise}$.  By examining equation~\eqref{EqnDefnDelfun} from
Lemma~\ref{LemUpperBer}, we see that by choosing $\topdeg$
sufficiently large, we can make $\delfun{\ldpcthresh}$ arbitrarily
close to $\myhalf$, and hence $\rent{\channoise}{\delfun{\ldpcthresh}
\ast \channoise}$ arbitrarily close to $1-\binent{\channoise}$.  More
precisely, let us choose $\topdeg$ large enough to guarantee that
$\rent{\channoise}{\delfun{\ldpcthresh} \ast \channoise} <
(1-\epsilon) \, (1 - \binent{\channoise})$, where \mbox{$\epsilon =
\frac{\ratetot \,(1-\binent{\lowtay})}{1-\binent{\channoise}}$.}  With
this choice, we have, for all $\mywei \in [\ldpcthresh, \lowtay]$, the
sequence of inequalities
\begin{eqnarray*}
\ErrFunTil(\mywei) & \leq & \ratecom \binent{\lowtay} -
\rent{\channoise}{\delfun{\ldpcthresh} \ast \channoise} \\
& < & \ratecom \binent{\lowtay} - \big[(1-\binent{\channoise}) -
\ratetot(1-\binent{\lowtay})\big] \\
&= & \ratecom - (1-\binent{\channoise}) \, < \, 0,
\end{eqnarray*}
which completes the proof.

\end{proof}


\section{Discussion}
\label{SecDiscussion}

In this paper, we established that it is possible to achieve both the
rate-distortion bound for symmetric Bernoulli sources and the channel
capacity for the binary symmetric channel using codes with bounded
graphical complexity.  More specifically, we have established that
there exist low-density generator matrix (LDGM) codes and low-density
parity check (LDPC) codes with finite degrees that, when suitably
compounded to form a new code, are optimal for both source and channel
coding.  To the best of our knowledge, this is the first demonstration
of classes of codes with bounded graphical complexity that are optimal
as source and channel codes simultaneously.  We also demonstrated that
this compound construction has a naturally nested structure that can
be exploited to achieve the Wyner-Ziv bound~\cite{WynerZiv76} for
lossy compression of binary data with side information, as well as the
Gelfand-Pinsker bound~\cite{Gelfand80} for channel coding with side
information.

Since the analysis of this paper assumed optimal decoding and
encoding, the natural next step is the development and analysis of
computationally efficient algorithms for encoding and decoding.
Encouragingly, the bounded graphical complexity of our proposed codes
ensures that they will, with high probability, have high girth and
good expansion, thus rendering them well-suited to message-passing and
other efficient decoding procedures.  For pure channel coding,
previous work~\cite{Etesami06,Pfister05,Shokrollahi06} has analyzed
the performance of belief propagation when applied to various types of
compound codes, similar to those analyzed in this paper.  On the other
hand, for pure lossy source coding, our own past work~\cite{WaiMan05}
provides empirical demonstration of the feasibility of modified
message-passing schemes for decoding of standard LDGM codes.  It
remains to extend both these techniques and their analysis to more
general joint source/channel coding problems, and the compound
constructions analyzed in this paper.


\subsection*{Acknowledgements}  The work of MJW was supported by 
National Science Foundation grant CAREER-CCF-0545862, a grant from
Microsoft Corporation, and an Alfred P. Sloan Foundation Fellowship.


\appendix

\section{Basic property of LDGM codes}
\label{AppUpperBer}

For a given weight $\mywei \in (0,1)$, suppose that we enforce that
the information sequence $\infobit \in \{0,1\}^\numinfobit$ has
exactly $\lceil \mywei \numinfobit \rceil$ ones.  Conditioned on this
event, we can then consider the set of all codewords $\Codebit(\mywei)
\in \{0,1\}^\numcodebit$, where we randomize over low-density
generator matrices $\Genmat$ chosen as in step (a) above.  Note for
any fixed code, $\Codebit(\mywei)$ is simply some codeword, but
becomes a random variable when we imagine choosing the generator
matrix $\Genmat$ randomly.  The following lemma characterizes this
distribution as a function of the weight $\mywei$ and the LDGM top
degree $\topdeg$:
\blems
\label{LemUpperBer}
Given a binary vector $\infobit \in \{0,1\}^\numinfobit$ with a
fraction $\mywei$ of ones, the distribution of the random LDGM
codeword $\Codebit(\mywei)$ induced by $\infobit$ is i.i.d. Bernoulli
with parameter $\delfun{\mywei} = \onehalf \, \biggr[ 1 - (1 - 2 \,
\mywei)^{\topdeg} \biggr]$.
\elems
\begin{proof}
Given a fixed sequence $\infobit \in \{0,1\}^\numinfobit$ with a
fraction $\mywei$ ones, the random codeword bit $\Codebit_i(\mywei)$
at bit $i$ is formed by connecting $\topdeg$ edges to the set of
information bits.\footnote{In principle, our procedure allows two
different edges to choose the same information bit, but the
probability of such double-edges is asymptotically negligible.}  Each
edge acts as an i.i.d. Bernoulli variable with parameter $\mywei$, so
that we can write
\begin{eqnarray}
\Codebit_i(\mywei) & = &  V_1 \oplus V_2 \oplus \ldots \oplus V_{\topdeg},
\end{eqnarray}
where each $V_k \sim \myber(\mywei)$ is independent and identically
distributed.  A straightforward calculation using z-transforms
(see~\cite{Gallager63}) or Fourier transforms over $GF(2)$ yields that
$\Codebit_i(\mywei)$ is Bernoulli with parameter $\delfun{\mywei}$ as
defined.
\end{proof}

\section{Bounds on binomial coefficients}
\label{AppBinCoeff}

The following bounds on binomial coefficients are standard (see
Chap. 12,~\cite{Cover}):
\begin{equation}
\binent{\frac{k}{n}} - \frac{\log (n+1)}{n} \; \leq \; \frac{1}{n} \log {n
\choose k} \; \leq \; \binent{\frac{k}{n}}.
\end{equation}
Here, for $\alpha \in (0,1)$, the quantity $h(\alpha) \mydefn -\alpha
\log \alpha -(1-\alpha) \log(1-\alpha)$ is the binomial entropy
function.


\section{Proof of Lemma~\ref{LemSecondMomentSimple}}
\label{AppSecondMomentSimple}

First, by the definition of $\Numc(\distor)$, we have
\begin{eqnarray*}
\Exs[\Numc^2(\distor)] & = & \Exs \left[\sum_{i=1}^{\numcw-1}
  \sum_{j=0}^{\numcw-1} \cwind{i}(\distor) \cwind{j}(\distor) \right]
  \\
& = & \Exs[\Numc] + \sum_{i=0}^{\numcw-1} \sum_{j \neq i}
\mprob[\cwind{i}(\distor) = 1, \, \cwind{i}(\distor) = 1].
\end{eqnarray*}
To simplify the second term on the RHS, we first note that for any
i.i.d Bernoulli($\myhalf$) sequence $\Ssca \in \{0,1\}^\numbit$ and
any codeword $\cw{j}$, the binary sequence $\Ssca' \mydefn \Ssca
\oplus \cw{j}$ is also i.i.d. Bernoulli($\myhalf$).  Consequently, for
each pair $i \neq j$, we have
\begin{eqnarray*}
\mprob \left[\cwind{i}(\distor) = 1, \, \cwind{j}(\distor) = 1 \right]
& = & \mprob \left[\|\cw{i} \oplus \Ssca\|_1 \leq \distor \numbit,
\|\cw{j} \oplus \Ssca \|_1\leq \distor \numbit \right] \\
& = & \mprob \left[\|\cw{i} \oplus \Ssca'\|_1 \leq \distor \numbit,
\|\cw{j} \oplus \Ssca' \|_1 \leq \distor \numbit \right ] \\
& = & \mprob \left[\|\cw{i} \oplus \cw{j} \oplus \Ssca \|_1 \leq
\distor \numbit, \|\Ssca \|_1 \leq \distor \numbit \right].
\end{eqnarray*}
Note that for each $j \neq i$, the vector $\cw{i} \oplus \cw{j}$ is a
non-zero codeword.  For each fixed $i$, summing over $j \neq i$ can be
recast as summing over all non-zero codewords, so that
\begin{eqnarray*}
\sum_{i \neq j} \mprob \left[\cwind{i}(\distor) = 1, \,
\cwind{j}(\distor) = 1 \right] & = & \sum_{i=0}^{\numcw-1} \sum_{j
\neq i} \mprob \left[\|\cw{i} \oplus \cw{j} \oplus \Ssca \|_1 \leq
\distor \numbit, \|\Ssca \|_1 \leq \distor \numbit \right] \\
& = & \sum_{i=0}^{\numcw -1} \sum_{k \neq 0} \mprob \left[\|\cw{k}
\oplus \Ssca \|_1 \leq \distor \numbit, \|\Ssca \|_1 \leq \distor
\numbit \right] \\
& = & 2^{\numbit \rate} \, \sum_{k \neq 0} \mprob \left[ \|\cw{k}
\oplus \Ssca \|_1 \leq \distor \numbit, \|\Ssca \|_1 \leq \distor
\numbit\right] \\
& = & 2^{\numbit \rate} \mprob \left[ \cwind{0}(\distor) = 1 \right]
\; \sum_{k \neq 0} \mprob \left[\cwind{k}(\distor) = 1 \, \mid \;
\cwind{0}(\distor) = 1\right] \\
& = & \Exs[\Numc] \; \sum_{k \neq 0} \mprob \left[\cwind{k}(\distor) =
1 \, \mid \; \cwind{0}(\distor) \right]
\end{eqnarray*}
thus establishing the claim.


\section{Proof of Lemma~\ref{LemQprob}}
\label{AppQprob}

%
We reformulate the probability $\Qprob(\mywei, \distor)$ as follows.
Recall that $\Qprob$ involves conditioning the source sequence $\Svar$
on the event $\|\Svar\|_1 \leq \distor \numcodebit$.  Accordingly, we
define a discrete variable $\rcountvar$ with distribution
\begin{eqnarray*}
\Prob(\rcountvar = \rval) & = & \frac{{\nbit \choose
\rval}}{\sum_{s=0}^{\distor \nbit} {\nbit \choose s}} \qquad \mbox{for
$\rval = 0,1, \ldots, \distor \nbit$},
\end{eqnarray*}
representing the (random) number of $1$s in the source sequence
$\Svar$.  Let $\rvaradd_i$ and $\rvarplain_j$ denote Bernoulli random
variables with parameters $1-\IndBer{\mywei}$ and $\IndBer{\mywei}$
respectively.  With this set-up, conditioned on codeword $j$ having a
fraction $\mywei \nbit$ ones, the quantity $\Qprob(\mywei, \distor)$
is equivalent to the probability that the random variable
\begin{eqnarray}
\label{EqnRandomFormulate}
\rtotvar & \mydefn & \begin{cases} \sum_{i=1}^\rcountvar \rvaradd_j +
\sum_{j=1}^{n - \rcountvar} \rvarplain_j  & \mbox{if $\rcountvar \geq 1$} \\
\sum_{j=1}^{n} \rvarplain_j & \mbox{if $\rcountvar =0$}
		     \end{cases}
\end{eqnarray}
is less than $\distor \nbit$.  To bound this probability, we use a
Chernoff bound in the form
\begin{eqnarray}
\label{EqnChernoff}
\frac{1}{\nbit} \log \Prob[\rtotvar\leq \distor \nbit] & \leq &
\inf_{\lambda < 0} \left( \frac{1}{\nbit} \log
\MomGen{\rtotvar}(\lambda) - \lambda \distor \right).
\end{eqnarray}
We begin by computing the moment generating function $\MomGen{\rtotvar}$.
Taking conditional expectations and using independence, we have
\begin{eqnarray*}
\MomGen{\rtotvar}(\lambda) & = & \sum_{\rval = 0}^{\distor \nbit}
 \Prob[\rcountvar = \rval] \; \left[\MomGen{\rvaradd}(\lambda)
 \right]^\rval \left[\MomGen{\rvarplain}(\lambda)\right]^{\nbit -
 \rval}.
\end{eqnarray*}
Here the cumulant generating functions have the form
\begin{subequations}
\begin{eqnarray}
\log \MomGen{\rvaradd}(\lambda) & = & \log \left[(1-\delta) e^\lambda
+ \delta \right], \quad \mbox{and} \\
\log \MomGen{\rvarplain}(\lambda) & = & \log \left[ (1-\delta) +
\delta e^\lambda \right],
\end{eqnarray}
\end{subequations}
where we have used (and will continue to use) $\delta$ as a shorthand
for $\IndBer{\mywei}$.

Of interest to us is the exponential behavior of this expression in
$\nbit$.  Using the standard entropy approximations to the binomial
coefficient (see Appendix~\ref{AppBinCoeff}), we can bound
$\MomGen{\rtotvar}(\lambda)$ as
\begin{eqnarray}
\label{EqnGbound}
&& f(\topbit) \; \sum_{\rval=0}^{\distor \nbit} \underbrace{\exp \big[
\nbit \big\{ \binent{\frac{\rval}{\nbit}} - \binent{\distor} +
\frac{\rval}{\nbit} \log \MomGen{\rvaradd}(\lambda) +
\left(1-\frac{\rval}{\nbit}\right) \log \MomGen{\rvarplain}(\lambda)
\big \} \big] },  \\
&& \qquad \qquad \qquad \qquad \qquad \qquad \qquad \qquad \qquad
\Tmpfun(t) \nonumber
\end{eqnarray}
where $f(\topbit)$ denotes a generic polynomial factor.  Further
analyzing this sum, we have
\begin{eqnarray}
\frac{1}{\nbit} \log \sum_{\rval=0}^{\distor \nbit} \Tmpfun(\rval) &
\leq & \frac{1}{\nbit} \max_{0 \leq \rval \leq \distor \nbit} \log
\Tmpfun(\rval) + \frac{\log f(\topbit)}{\topbit} + \frac{\log (\nbit
\distor)}{\nbit} \nonumber \\
& = & \max_{0 \leq \rval \leq \distor \nbit} \left\{
\binent{\frac{\rval}{\nbit}} - \binent{\distor} + \frac{\rval}{\nbit}
\log \MomGen{\rvaradd}(\lambda) + \left(1-\frac{\rval}{\nbit}\right)
\log \MomGen{\rvarplain}(\lambda) \right \} + o(1) \nonumber \\
& \leq & \max_{\tmpq \in [0,\distor]} \left\{ \binent{\tmpq} -
\binent{\distor} + \tmpq \log \MomGen{\rvaradd}(\lambda) +
\left(1-\tmpq\right) \log \MomGen{\rvarplain}(\lambda) \right \} +
o(1). \nonumber
\end{eqnarray}
Combining this upper bound on $\frac{1}{\nbit} \log
\MomGen{\rtotvar}(\lambda)$ with the Chernoff
bound~\eqref{EqnChernoff} yields that
\begin{eqnarray}
\label{EqnFinalUpper}
\frac{1}{\nbit} \log \Prob[\rtotvar\leq \distor \nbit] & \leq &
 \inf_{\lambda < 0} \max_{\tmpq \in [0, \distor]} \InterFunc(\tmpq,
 \lambda; \delta) + o(1)
\end{eqnarray}
where the function $\InterFunc$ takes the form
\begin{eqnarray}
\label{EqnDefnInterFunc}
\InterFunc(\tmpq, \lambda; \delta) & \mydefn & \binent{\tmpq} -
\binent{\distor} + \tmpq \log \MomGen{\rvaradd}(\lambda) +
\left(1-\tmpq\right) \log \MomGen{\rvarplain}(\lambda) - \lambda
\distor.
\end{eqnarray}

Finally, we establish that the solution $(\ustar, \lambda^*)$ to the
min-max saddle point problem~\eqref{EqnFinalUpper} is unique, and
specified by $\ustar = \distor$ and $\lamstar$ as in
Lemma~\ref{LemQprob}.  First of all, observe that for any $\delta \in
(0,1)$, the function $\InterFunc$ is continuous, strictly concave in
$\tmpq$ and strictly convex in $\lambda$.  (The strict concavity
follows since $\binent{\tmpq}$ is strictly concave with the remaining
terms linear; the strict convexity follows since cumulant generating
functions are strictly convex.) Therefore, for any fixed $\lambda <
0$, the maximum over $\tmpq \in [0, \distor]$ is always achieved.  On
the other hand, for any $\distor > 0$, $\tmpq \in [0, \distor]$ and
$\delta \in (0,1)$, we have $\InterFunc(\tmpq; \lambda; \tmpvar)
\rightarrow +\infty$ as $\lambda \rightarrow -\infty$, so that the
infimum is either achieved at some $\lamstar < 0$, or at $\lamstar =
0$.  We show below that it is always achieved at an interior point
$\lamstar < 0$.  Thus far, using standard saddle point
theory~\cite{Hiriart1}, we have established the existence and
uniqueness of the saddle point solution $(\ustar, \lambda^*)$.

To verify the fixed point conditions, we compute partial derivatives in
order to find the optimum.  First, considering $\tmpq$, we compute
\begin{eqnarray*}
\frac{\partial \InterFunc}{\partial \tmpq}(\tmpq, \lambda; \delta) & =
& \log \frac{1-\tmpq}{\tmpq} + \log \MomGen{\rvaradd}(\lambda) -\log
\MomGen{\rvarplain}(\lambda) \\
& = & \log \frac{1-\tmpq}{\tmpq} +\log \left[(1-\delta) e^\lambda +
\delta \right] - \log \left[ (1-\delta) + \delta e^\lambda \right].
\end{eqnarray*}
Solving the equation $\frac{\partial \InterFunc}{\partial
\tmpq}(\tmpq, \lambda; \delta) = 0$ yields
\begin{eqnarray}
u' & = & \frac{\exp(\lambda)}{1+\exp(\lambda)} \distor +
\frac{1}{1+\exp(\lambda)} (1-\distor) \; \geq \; 0.
\end{eqnarray}
Since $\distor \leq \myhalf$, a bit of algebra shows that $u' \geq D$
for all choices of $\lambda$.  Since the maximization is constrained
to $[0, \distor]$, the optimum is always attained at $u^* = D$.

Turning now to the minimization over $\lambda$, we compute the partial
derivative to find
\begin{eqnarray*}
\frac{\partial \InterFunc}{\partial \lambda}(\tmpq, \lambda; \delta) &
= & \tmpq \frac{(1-\delta) \exp(\lambda)}{(1-\delta) \exp(\lambda) +
\delta} + (1-\tmpq) \frac{\delta \exp(\lambda)}{(1-\delta) + \delta
\exp(\lambda)} - \distor.
\end{eqnarray*}
Setting this partial derivative to zero yields a quadratic equation in
$\exp(\lambda)$ with coefficients
\begin{subequations}
\begin{eqnarray}
\label{EqnQuadRoots}
\acoeff & = & \delta \, (1-\delta) \, (1-\distor) \\
\bcoeff & = & \tmpq (1-\delta)^2 + (1-\tmpq) \delta^2 - \distor
\left[\delta^2 + (1-\delta)^2 \right]. \\
\ccoeff & = & - \distor \delta (1-\delta).
\end{eqnarray}
\end{subequations}
The unique positive root $\rho^*$ of this quadratic equation is given
by
\begin{eqnarray}
\label{EqnDefnRhoStar}
\rho^*(\delta, \distor, \tmpq) & \mydefn & \frac{1}{2 \acoeff}
\left[-\bcoeff + \sqrt{\bcoeff^2 -4 \acoeff \ccoeff} \right].
\end{eqnarray}

It remains to show that $\rho^* \leq 1$, so that $\lamstar \mydefn
\log \rho^* < 0$.  A bit of algebra (using the fact $\acoeff \geq 0$)
shows that $\rho^* < 1$ if and only if $\acoeff + \bcoeff + \ccoeff >
0$.  We then note that at the optimal $\tmpq^* = \distor$, we have
$\bcoeff = (1-2\distor) \delta^2$, whence
\begin{eqnarray*}
\acoeff + \bcoeff + \ccoeff & = & \delta \, (1-\delta) \, (1-\distor)
+ (1-2 \distor) \delta^2 - \distor \delta (1-\delta) \\
& = & (1-2 \distor) \, \delta \; > \; 0
\end{eqnarray*}
since $\distor < \myhalf$ and $\delta > 0$.  Hence, the optimal
solution is $\lambda^* \mydefn \log \rho^* < 0$, as specified in the
lemma statement.

\section{Proof of Lemma~\ref{LemFbound}}
\label{AppFbound}

A straightforward calculation yields that 
\[
G(\myhalf) = \KeyFunc(\IndBer{\myhalf}; \distor) = \KeyFunc(\myhalf;
\distor) = -\left(1-\binent{\distor} \right) 
\] 
as claimed.  Turning next to the derivatives, we note that by
inspection, the solution $\lamstar(\tmpvar)$ defined in
Lemma~\ref{LemQprob} is twice continuously differentiable as a
function of $\tmpvar$.  Consequently, the function $\KeyFunc(t,
\distor)$ is twice continuously differentiable in $\tmpvar$.
Moreover, the function $\IndBer{\mywei}$ is twice continuously
differentiable in $\mywei$.  Overall, we conclude that $G(\mywei) =
\KeyFunc(\IndBer{\mywei}; \distor)$ is twice continuously
differentiable in $\mywei$, and that we can obtain derivatives via
chain rule.  Computing the first derivative, we have
\begin{eqnarray*}
G'(\myhalf) & = & \delta'(\myhalf) \, \KeyFunc'(\IndBer{\myhalf};
\distor) \; = \; 0
\end{eqnarray*}
since $\delta'(\mywei) = - \topdeg \, (1-2\mywei)^{\topdeg-1}$, which
reduces to zero at $\mywei = \myhalf$.  Turning to the second
derivative, we have
\begin{eqnarray*}
G''(\myhalf) & = & \delta''(\myhalf) \, \KeyFunc'(\IndBer{\myhalf};
\distor) + \left(\delta'(\myhalf)\right)^2
\KeyFunc''(\IndBer{\myhalf}; \distor) \; = \; \delta''(\myhalf) \,
\KeyFunc'(\IndBer{\myhalf}; \distor).
\end{eqnarray*}
We again compute $\delta''(\mywei) = 2 \topdeg \, (\topdeg-1)
(1-2\mywei)^{\topdeg-2}$, which again reduces to zero at $\mywei =
\myhalf$ since $\topdeg \geq 4$ by assumption.


\section{Regular LDPC codes are sufficient}
\label{AppLDPCSuffice}

Consider a regular $(\vdeg, \lowcdeg)$ code from the standard Gallager
LDPC ensemble.  In order to complete the proof of
Theorem~\ref{ThmSource}, we need to show for suitable choices of
degree $(\vdeg, \lowcdeg)$, the average weight enumerator of these
codes can be suitably bounded, as in equation~\eqref{EqnLDPCAss}, by a
function $\BouWtEnum$ that satisfies the conditions specified in
Section~\ref{SecFiniteDegrees}.

\begin{figure}[h]
\begin{center}
\widgraph{0.5\textwidth}{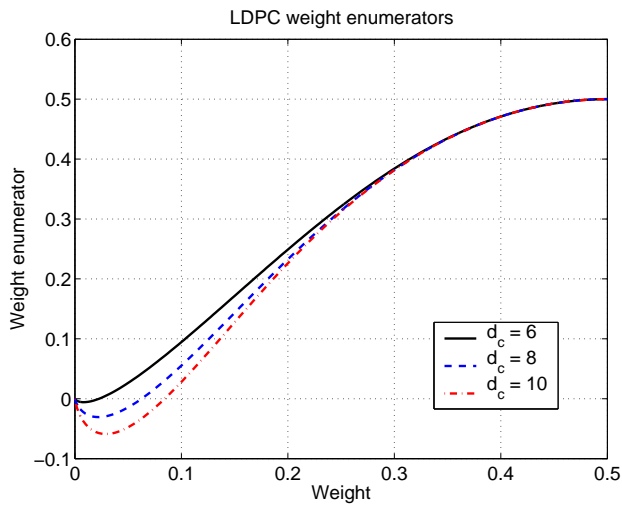}
\caption{Plots of LDPC weight enumerators for codes of rate $\rateldpc = 0.5$,
and check degrees $\lowcdeg \in \{6, 8, 10 \}$.}
\label{FigGallager}
\end{center}
\end{figure}

It can be shown~\cite{Gallager63,Litsyn02} that for even degrees
$\lowcdeg$, the average weight enumerator of the regular Gallager
ensemble, for any block length $\numinfobit$, satisfies the bound
\begin{eqnarray*}
\frac{1}{\numinfobit} \log \AvWtEnum{\numinfobit}(\mywei) & = &
\BouWtEnum(\mywei; \vdeg, \lowcdeg) + o(1).
\end{eqnarray*}
The function $\BouWtEnum$ in this relation is defined for $\mywei \in
[0, \myhalf]$ as
\begin{eqnarray}
\label{EqnDefnBouWtEnum}
\BouWtEnum(\mywei; \vdeg, \lowcdeg) & \mydefn & (1-\vdeg)
\binent{\mywei} - (1-\rateldpc) + \vdeg \inf_{\mylam \leq 0 } \left\{
\frac{1}{\lowcdeg} \log \left((1+e^\mylam)^{\lowcdeg} +
(1-e^\mylam)^{\lowcdeg} \right) - \mywei \mylam \right\}, \qquad
\end{eqnarray}
and by $\BouWtEnum(\mywei) = \BouWtEnum(\mywei - \myhalf)$ for $\mywei
\in [\myhalf, 1]$. Given that the minimization
problem~\eqref{EqnDefnBouWtEnum} is strictly convex, a straightforward
calculation of the derivative shows the optimum is achieved at
$\mylam^*$, where $\mylam^* \leq 0$ is the unique solution of the
equation
\begin{eqnarray}
\label{EqnDefnLamstar}
 e^\mylam \frac{(1+e^\mylam)^{\lowcdeg-1} - (1-e^\mylam)^{\lowcdeg-1}}
{(1+e^\mylam)^{\lowcdeg} + (1-e^\mylam)^{\lowcdeg}} & = & \mywei.
\end{eqnarray}
Some numerical computation for $\rateldpc = 0.5$ and different choices
$(\vdeg, \lowcdeg)$ yields the curves shown in~\figref{FigGallager}.

We now show that for suitable choices of degree $(\vdeg, \lowcdeg)$,
the function $\BouWtEnum$ defined in equation~\eqref{EqnDefnBouWtEnum}
satisfies the four assumptions specified in
Section~\ref{SecFiniteDegrees}.  First, for even degrees $\lowcdeg$,
the function $\BouWtEnum$ is symmetric about $\mywei = \myhalf$, so
that assumption (A1) holds.  Secondly, we have $\BouWtEnum(\mywei)
\leq \rateldpc$, and moreover, for $\mywei = \myhalf$, the optimal
$\mylam^*(\myhalf) = 0$, so that $\BouWtEnum(\myhalf) = \rateldpc$,
and assumption (A3) is satisfied.  Next, it is known from the work of
Gallager~\cite{Gallager63}, and moreover is clear from the plots
in~\figref{FigGallager}, that LDPC codes with $\vdeg > 2$ have linear
minimum distance, so that assumption (A4) holds.

The final condition to verify is assumption (A2), concerning the
differentiability of $\BouWtEnum$.  We summarize this claim in the
following:
\blems
The function $\BouWtEnum$ is twice continuously differentiable on
$(0,1)$, and in particular we have
\begin{equation}
\BouWtEnum'(\myhalf) \; = \; 0, \qquad \mbox{and} \qquad
\BouWtEnum''(\myhalf) \; < \; 0.
\end{equation}
\begin{proof}
Note that for each fixed $\mywei \in (0,1)$, the function
\begin{eqnarray*}
f(\mylam) & = & \frac{1}{\lowcdeg} \log \left((1+e^\mylam)^{\lowcdeg}
+ (1-e^\mylam)^{\lowcdeg} \right) \; = \; \frac{1}{\lowcdeg} \log
\left((e^{-\mylam}+1)^{\lowcdeg} + (e^{-\mylam}-1)^{\lowcdeg} \right)
+ \mylam
\end{eqnarray*}
is strictly convex and twice continuously differentiable as a function
of $\lambda$.  Moreover, the function \mbox{$f^*(\mywei) \mydefn
\inf_{\lambda \leq 0} \ \{ f(\mylam) - \mylam \mywei \}$} corresponds
to the conjugate dual~\cite{Hiriart1} of $f(\lambda) +
\mathbb{I}_{\leq 0}(\lambda)$.  Since the optimum is uniquely attained
for each $\mywei \in (0,1)$, an application of Danskin's
theorem~\cite{Bertsekas_nonlin} yields that $f^*$ is differentiable
with $\frac{d}{d\mywei} f^*(\mywei) = -\mylam^*(\mywei)$, where
$\mylam^*$ is defined by equation~\eqref{EqnDefnLamstar}.  Putting
together the pieces, we have $\BouWtEnum'(\mywei) = (1-\vdeg)
h'(\mywei) - \vdeg \mylam^*(\mywei)$.  Evaluating at $\mywei =
\myhalf$ yields $\BouWtEnum'(\myhalf) = 0 - \vdeg \mylam^*(0) = 0$ as
claimed.

We now claim that $\mylam^*(\mywei)$ is differentiable.  Indeed, let
us write the defining relation~\eqref{EqnDefnLamstar} for
$\mylam^*(\mywei)$ as $F(\mylam, \mywei) = 0$ where $F(\mylam, \mywei)
\mydefn f'(\mylam) - \mywei$.  Note that $F$ is twice continuously
differentiable in both $\mylam$ and $\mywei$; moreover,
$\frac{\partial F}{\partial \mylam}$ exists for all $\lambda \leq 0$
and $\mywei$, and satisfies $\frac{\partial F}{\partial
\mylam}(\mylam, \mywei) = f''(\mylam) > 0$ by the strict convexity of
$f$.  Hence, applying the implicit function
theorem~\cite{Bertsekas_nonlin} yields that $\mylam^*(\mywei)$ is
differentiable, and moreover that $\frac{d \mylam^*}{d \mywei}(\mywei)
= 1/f''(\mylam^*(\mywei))$.  Hence, combined with our earlier
calculation of $\BouWtEnum'$, we conclude that $\BouWtEnum''(\mywei) =
(1-\vdeg) h''(\mywei) - \vdeg \frac{1}{f''(\mylam(\mywei))}$.  Our
final step is to compute the second derivative $f''$.  In order to do
so, it is convenient to define $g = \log f'$, and exploit the relation
$g' f' = f''$.  By definition, we have
\begin{eqnarray*}
g(\mylam) & = & \mylam + \log \left[(1+e^\mylam)^{\lowcdeg-1} -
(1-e^\mylam)^{\lowcdeg-1} \right] - \log \left[(1+e^\mylam)^{\lowcdeg}
+ (1-e^\mylam)^{\lowcdeg} \right]
\end{eqnarray*}
whence
\begin{eqnarray*}
g'(\mylam) & = & 1 + e^\mylam (\lowcdeg-1)
\frac{(1+e^\mylam)^{\lowcdeg-2} +
(1-e^\mylam)^{\lowcdeg-2}}{(1+e^\mylam)^{\lowcdeg-1} -
(1-e^\mylam)^{\lowcdeg-1}} - e^\mylam \lowcdeg
\frac{(1+e^\mylam)^{\lowcdeg-1} -
(1-e^\mylam)^{\lowcdeg-1}}{(1+e^\mylam)^{\lowcdeg} +
(1-e^\mylam)^{\lowcdeg}}
\end{eqnarray*}
Evaluating at $\mywei = \myhalf$ corresponds to $\mylam(0) = 0$, so
that
\begin{eqnarray*}
f''(\mylam(\myhalf)) & = & f'(0) \, g'(0) \; = \; \myhalf \; \left[1 +
  (\lowcdeg-1) \frac{2^{\lowcdeg-2}}{2^{\lowcdeg-1}} - \lowcdeg
 \frac{2^{\lowcdeg-1}}{2^{\lowcdeg}} \right] \; = \; \frac{1}{4}.
\end{eqnarray*}
Consequently, combining all of the pieces, we have
\begin{equation*}
\BouWtEnum''(\mywei) = (1-\vdeg) h''(\myhalf) - \vdeg
\frac{1}{f''(\mylam(\myhalf))} \; = \; \frac{\vdeg-1}{4} - 4 \vdeg < 0
\end{equation*}
as claimed.

\end{proof}

\elems

\section{Derivatives of $\ErrFunTil$}
\label{AppErrFunTilDeriv}

Here we calculate the first and second derivatives of the function
$\ErrFunTil$ defined in equation~\eqref{EqnDefnErrFunTil}.  The first
derivative takes the form
\begin{eqnarray*}
\ErrFunTil'(\weight) & = & \ratecom \log \frac{1-\weight}{\weight} +
\channoise
\frac{\delta'(\weight;\topdeg)}{\inducedWeight{\weight;\topdeg}} -
(1-\channoise)
\frac{\delta'(\weight;\topdeg)}{1-\inducedWeight{\weight;\topdeg}}
\end{eqnarray*}
where $\delta'(\weight; \topdeg) = \topdeg (1-2 \weight)^{\topdeg-1}$.
Since $\delta'(\myhalf;\topdeg) = 0$, we have $\ErrFunTil'(\myhalf) =
0$ as claimed.  Second, using chain rule, we calculate
\begin{multline*}
\ErrFunTil''(\weight) = -\ratecom \big[\frac{1}{1-\weight}
+\frac{1}{\weight} \big] + \channoise \frac{\delta''(\weight;\topdeg)
\inducedWeight{\weight;\topdeg} -
[\delta'(\weight;\topdeg)]^2}{[\inducedWeight{\weight;\topdeg}]^2}\\
 - (1-\channoise) \frac{\delta''(\weight;\topdeg)
\big[1-\inducedWeight{\weight;\topdeg}\big] +
[\delta'(\weight;\topdeg)]^2}{[1-\inducedWeight{\weight;\topdeg}]^2}
\end{multline*}
and $\delta''(\weight;\topdeg) = -\topdeg \, (\topdeg-1) \,
(1-2\weight)^{\topdeg-2}$.  Now for $\topdeg > 2$, we have
$\delta''(\myhalf) = 0$, so that $\ErrFunTil''(\myhalf) = - 4 \ratecom
< 0$ as claimed.

\end{onehalfspace}


\bibliographystyle{latex8} 

\end{document}